\documentclass[sigconf,natbib=true,screen=true]{acmart}
\usepackage{type1cm}     % type1 computer modern font
\usepackage{graphicx}     % advanced figures
\usepackage{xspace}     % fix space in macros
\usepackage{booktabs}     % nicer tables
\usepackage{multi row}     % multi rows for tables
\usepackage[font={bf}, tableposition=top]{caption}     % captions on top for tables
\usepackage{bold-extra}     % bold + {small capital, italic}
\usepackage[vlined,linesnumbered,ruled,noend]{algorithm2e}     % algorithms
\usepackage{microtype}    % compress text
\usepackage{units}     % nicer slanted fractions
\usepackage{mathtools}     % amsmath++
\usepackage{amssymb}     % math symbols
\usepackage{todonotes}
\usepackage[utf8]{inputenc}
\usepackage{xcolor}
\usepackage{rotating}
\usepackage{subfig} % multiple sub-figures
\usepackage{placeins}

\usepackage{enumitem}

\newlist{steps}{enumerate}{1}
\setlist[steps, 1]{label = RQ\arabic*:}
\usepackage{flushend}
\usepackage{xcolor}
\usepackage{graphicx}
\usepackage{balance}

\newcommand\Tf{\rule{0pt}{2.8ex}}

\setcopyright{acmcopyright}

\title{Facebook Ads as a Demographic Tool to Measure the Urban-Rural Divide}

\author{Daniele Rama}
\orcid{0000-0001-9419-7315}
\affiliation{%
  \institution{University of Turin \& ISI Foundation}
  \city{Turin}
  \country{Italy}
}
\email{daniele.rama@unito.it}

\author{Yelena Mejova}
\orcid{0000-0001-5560-4109}
\affiliation{%
  \institution{ISI Foundation}
  \city{Turin}
  \country{Italy}
}
\email{yelenamejova@acm.org}

\author{Michele Tizzoni}
\orcid{0000-0001-7246-2341}
\affiliation{%
  \institution{ISI Foundation}
  \city{Turin}
  \country{Italy}
}
\email{michele.tizzoni@isi.it}

\author{Kyriaki Kalimeri}
\orcid{0000-0001-8068-5916}
\affiliation{%
  \institution{ISI Foundation}
  \city{Turin}
  \country{Italy}
}
\email{kkalimeri@acm.org}

\author{Ingmar Weber}
\orcid{0000-0003-4169-2579}
\affiliation{%
  \institution{Qatar Computing Research Institute}
  \city{Doha}
  \country{Qatar}
}
\email{ingmarweber@acm.org}

\begin{document}

\begin{abstract}
In the global move toward urbanization, making sure the people remaining in rural areas are not left behind in terms of development and policy considerations is a priority for governments worldwide.
However, it is increasingly challenging to track important statistics concerning this sparse, geographically dispersed population, resulting in a lack of reliable, up-to-date data.
In this study, we examine the usefulness of the Facebook Advertising platform, which offers a digital ``census'' of over two billions of its users, in measuring potential rural-urban inequalities.
We focus on Italy, a country where about 30\% of the population lives in rural areas. 
First, we show that the population statistics that Facebook produces suffer from instability across time and incomplete coverage of sparsely populated municipalities. 
To overcome such limitation, we propose an alternative methodology for estimating Facebook Ads audiences that nearly triples the coverage of the rural municipalities from 19\% to 55\% and makes feasible fine-grained sub-population analysis. 
Using official national census data, we evaluate our approach and confirm known significant urban-rural divides in terms of educational attainment and income.
Extending the analysis to Facebook-specific user ``interests'' and behaviors, we provide further insights on the divide, for instance, finding that rural areas show a higher interest in gambling.
Notably, we find that the most predictive features of income in rural areas differ from those for urban centres, suggesting researchers need to consider a broader range of attributes when examining rural wellbeing.
The findings of this study illustrate the necessity of improving existing tools and methodologies to include under-represented populations in digital demographic studies -- the failure to do so could result in misleading observations, conclusions, and most importantly, policies.
\end{abstract}

\copyrightyear{2020}
\acmYear{2020}
%\setcopyright{iw3c2w3}
\acmConference[WWW '20]{Proceedings of The Web Conference 2020}{April 20--24, 2020}{Taipei, Taiwan}
\acmBooktitle{Proceedings of The Web Conference 2020 (WWW '20), April 20--24, 2020, Taipei, Taiwan}
\acmPrice{}
\acmDOI{10.1145/3366423.3380118}
\acmISBN{978-1-4503-7023-3/20/04}




\maketitle

\section{Introduction}

In a rapidly urbanizing world, living in a rural community may present disadvantages from potentially residing far from a healthy food source (a ``food desert'') \cite{larsen2011welcome}, to lower wages \cite{young2013inequality}, to poorer health outcomes \cite{mainous1995comparison}.
Disadvantages continue when considering the study and measurement of these populations to motivate appropriate policies. 
Demographers have long acknowledged the instability of measures concerning rural populations due to sparsity \cite{murdock2008applied}, often methodologically mitigated by substituting statistics from larger areas with similar population characteristics, making trends observed in specific rural areas, in fact, synthetic \cite{Murdock2012}.

A possible solution may lie in a wealth of new digital data sources that has prompted a rise in recent research under the umbrella of ``Digital Demography'' \cite{alburez-gutierrez_zagheni_aref_gil-clavel_grow_negraia_2019}. 
The digitization of censuses \cite{thorvaldsenhandbook}, digital traces from online social networks \cite{kent2013spatial}, crowd-sourced data from participatory platforms \cite{potzschke2017migrant}, and internet-enabled devices \cite{roessger2017using} present new exciting opportunities for demographers by providing several advantages with respect to traditional sources~\cite{kalimeri2019predicting}.
Digital traces are generally accessible in high volume, often carry geographic information, and can be collected in real-time, allowing for the study of populations and their behaviors with unprecedented temporal and spatial granularity.

One such resource becoming popular in demography studies is Facebook's Advertising platform \cite{alburez-gutierrez_zagheni_aref_gil-clavel_grow_negraia_2019}, which provides advertisers with an estimate of a potential advertisement's reach, given the location, demographic, or behavioral constraints of the target audience\footnote{\url{https://developers.facebook.com/docs/marketing-api/buying-api/targeting}}.
Thus providing a digital ``census'' of its estimated 2.23 billion monthly active user base, this resource has prompted studies in tracking health conditions \cite{mejova2018online,rampazzo2018mater}, migration \cite{zagheni2017leveraging,spyratos2018migration}, crime \cite{fatehkia2019correlated}, and gender inequality \cite{garcia2018analyzing,mejova2018measuring,fatehkia2018using}. 
These studies illustrate the benefits of using massive social media platforms for the examination of, especially, difficult-to-reach populations such as women in India~\cite{mejova2018measuring}, or difficult-to-count ones such as migrants in Spain~\cite{spyratos2018migration}. 
Furthermore, the ability to track user ``interests'' beyond standard demographics, such as dietary habits, entertainment preferences, and technology use can extend the observations well beyond the standard demographic indicators \cite{mejova2018online}.
Thus, the application of this data to rural populations promises both a higher granularity and a richer palette of potential variables.
Nevertheless, given the nature of the service, its internal construction is a ``black box'', and a slew of biases, including Facebook user base self-selection, the algorithmic bias in extracting user attributes, and advertisement revenue incentives, must be taken into account by researchers~\cite{cesare2018promises}.

In this study, we examine Facebook Advertising as a resource for measuring the rural-urban divide in Italy, a country in which 30\% of its population is living in rural areas (76\% of total landmass)\footnote{\url{https://data.worldbank.org/indicator/SP.RUR.TOTL.ZS?locations=IT}}. In particular, we address three major research questions:

\begin{steps}
    \item How reliable are Facebook Advertising audience estimates across time and population density, especially considering rural or sparsely populated areas?
    \item How well do these estimates correspond to the official demographic figures?
    \item How can we enrich the current measurement of the urban-rural divide using Facebook Advertising?
\end{steps}

We contribute to the current state of the art in four ways.
First, we provide a systematic stability and spatial coverage analysis of the Facebook Advertising audience estimates for a wide set of user interests and behaviors, evidencing how both introduce particular disadvantages for rural municipalities.
Second, we propose an alternative use of the platform to overcome limitations of sparsely populated areas posed by the platform itself due to its rate limits.
Our methodology drastically improves coverage, especially of the rural municipalities, nearly tripling their coverage and making possible sub-dividing these populations for further analysis.
Third, evaluating our method on official national census data, we quantify urban-rural inequalities, regarding both standard demographic attributes such as socio-economic indicators, as well as novel behavioural estimates available through the Facebook platform.
We confirm a significant urban-rural divide, with Facebook users in urban areas having higher educational attainment and using higher-end cellphones (a proxy for income), while those in rural areas showing higher interest in gambling and Catholic church, doing more commuting, and using 3G or 4G networks instead of WiFi (pointing to a difference in internet access).
Finally, we model per capita income in rural, suburban, and urban areas using the Facebook indicators, and show that the variables most predictive of income in rural areas (commuting, use of 4G network) are different from those in urban areas (marital status, educational attainment, interest in fitness and wellness, etc.). 

Insights provided in this study are applicable to any re-purposing of digital resources for demographic research, pointing to the necessity of a cautious examination of any platform's coverage and stability before using information it provides. However, we also show that careful use of the platform's flexible querying process extends the usability of the data and allows for rich modeling of rural-urban inequality, augmenting analysis with Facebook users' many behavioral attributes.

\section{Related Work}

%Urban-rural inequalities in demographic and economic research
Assessing global trends in the urban-rural divide is an essential topic of research in economics and demography,
as almost everywhere in the world living standards of urban areas remain superior to those in rural areas \cite{sahn2003urban}.
Such divide can be observed across several different socioeconomic dimensions, ranging from per capita income to child mortality rates,  persisting even as countries develop into industrialized economies, as demonstrated by the cases of China and India \cite{sicular2007urban, saikia2013explaining}.
Historical trends in development, which tend to benefit those in already privileged positions, have resulted in, for example, technological disparities in terms of internet access and technological literacy -- trends which ongoing rural development policies attempt to address \cite{dimaggio2001digital,OECD2018}.
Even in OECD countries, persistent disparities between large metropolitan centres and rural areas are being recorded every year, as migration patterns intersect with other population attributes \cite{OECD2018migrant}.
According to a report of the World Bank, the urban-rural disparities increase quickly in early development until countries reach upper-middle-income levels \cite{rigg2009world}. Then, as countries grow, the gap becomes smaller, but convergence is usually slower.

%urban scaling theory
From a theoretical point of view, recent studies have investigated the structural advantage of cities in a wide range of output indicators, from patent production to personal income, through the robust framework of scaling \cite{bettencourt2007growth, depersin2018global}.
In particular, superlinear scaling of cities' growth has been explained as a consequence of increased social interactions with population density \cite{bettencourt2013origins} and by the process of selective migration of highly productive individuals into larger cities \cite{keuschnigg2019urban, young2013inequality}.
Understanding the mechanisms underlying the urban-rural divide remains an essential issue for policymaking, especially to make progress towards the ``Leaving no one behind'' pledge of the United Nations 2030 Agenda\footnote{\url{https://www.un.org/sustainabledevelopment/sustainable-development-goals/}}.

%digital demography: various sources of digital data
Leveraging on the immense amount of digital data produced daily, the field of Digital Demography emerged, addressing vital research questions of demographic research via innovative data sources.
These new sources of data are demonstrated to be particularly powerful in monitoring a series of demographic phenomena such as
birthrates~\cite{barclay2017long}, mortality ~\cite{baranowska2017effect}, unemployment~\cite{bonanomi2017understanding}, daily commuting ~\cite{beiro2016predicting}, international and internal migration~\cite{zagheni2014inferring}, but also modelling more complex socio-demographic issues such as psychological well-being and attitudes 
towards health~\cite{mejova2019effect,kalimeri2019human}.
Digital data are particularly useful in cases where official data are sparse, incomplete, or even impossible to obtain. 
For instance, Adler et al.~\cite{adler2019search} assessed the issue of suicide underreporting via query data, focusing on the Indian context, where social stigma and the only recent decriminalization of suicides, hampered the official agencies’ data collection.
%Digital demography and digital divide
Interestingly, digital sources can be used to investigate social inequalities. 
%Besides the well-known digital divide \cite{robinson2015digital,dimaggio2001digital}, 
In particular, social media data are proven to be useful in studying gender differences in access to technology \cite{mejova2018measuring,garcia2018analyzing} and parenting biases favouring male children mentions on social media~\cite{Sivak2039}.

%Facebook data
Among all online social networks, Facebook is the most popular one. In June 2018\footnote{\url{https://investor.fb.com/investor-news/press-release-details/2018/Facebook-Reports-Second-Quarter-2018-Results/default.aspx}}, Facebook reported 1.47 billion daily active users (DAUs) and 2.23 billion monthly active users (MAUs), with an increase of 11\% year-over-year. On average, three out of ten people used Facebook in 2018, and this estimate, considering the current trend, is destined to grow. 

Researchers have taken note of Facebook's massive user base, and, for instance, used it in the health domain for the recruitment of people affected by not-so-common conditions \cite{crosier2016using}, or a particular health behaviour \cite{carter2016beyond}. 
With the passing of the European General Data Protection Regulation (GDPR) more restrictions have been put on the processing and exploitation of personal data of individual users \cite{cabanas2018facebook}, such as political orientation, religious beliefs, ethnic origin, etc., due to the apparent privacy risks that may be derived from a malicious use of such type of information.

Yet, collecting individual user data is not necessary for the demographic study of populations. 
Instead, demographers recently began to use Facebook Advertising platform to gather statistics on select populations by querying for the number of users who an advertisement could reach.
In this fashion, it is possible to communicate a count of users matching specific socio-demographic characteristics at various geographic scales without revealing personal information of individual users.

Seminal works using this approach focused on the health domain \cite{mejova2018online,rampazzo2018mater}. For example, Ara\'ujo et al. \cite{araujo2017using} extracted data in 47 countries to track health conditions associated with lifestyle diseases. 
They showed that, within each country, Facebook data could provide insights into different trends of health awareness across demographic groups. 
Zagheni et al. \cite{zagheni2017leveraging} proposed the use of Facebook Advertising data to monitor stocks of migrants inside the US with promising results, paving the way for similar studies in the European context~\cite{spyratos2018migration,potzschke2017migrant}, but also more in depth studies on the assimilation of migrants in society \cite{dubois2018studying}.

Only recently, Facebook Advertising data were considered to examine social inequalities, and in particular gender inequalities in Internet access both at national \cite{garcia2018analyzing,fatehkia2018using} and sub-national levels \cite{mejova2018measuring}. 
In this direction, a study by Gil-Clavel and Zagheni \cite{gilclavel2019demographic} extended the analysis of the gender gap in Facebook adoption by adding the dimension of age.

All these works provided evidence that Facebook Ads data can indeed be used as a source of information for the study of digital disparities. However, little effort was made to understand the stability and representativity of such data across the several attributes available through the platform. 
As several studies pointed out, big observational data are not always representative of larger populations in the way that randomized surveys are~\cite{zagheni2015demographic,yildiz2017using,tufekci2014big,olteanu2019social}.
Coverage can also be an issue, as access to the internet is more restricted in low and middle-income countries, which can lead even to risks of re-identification \cite{cabanas2018facebook,salganik2019bit,de2015unique}.

This study contributes to the Digital Demography literature providing a thorough examination of the stability and coverage of Facebook Advertising data.
We particularly focus on behavioral signals related to social inequalities, especially those that may be more difficult to track officially, such as interests and hobbies.

\section{Data Collection \& Methods}

\subsection{Census data}
\label{sec:realworlddata}

%Define rural / urban
Italy is divided into 20 regions, which are subdivided into provinces, and then further into municipalities, or \emph{comuni}, which are the smallest administrative units. In order to differentiate between the urbanization within Italy, we consider the scale of municipalities, of which there were 7,978 as of February 20, 2019. Note that this number fluctuates, with municipalities merging, breaking up, and being redefined over time.

\begin{figure}[t]
    \centering
    \includegraphics[width=0.60\linewidth]{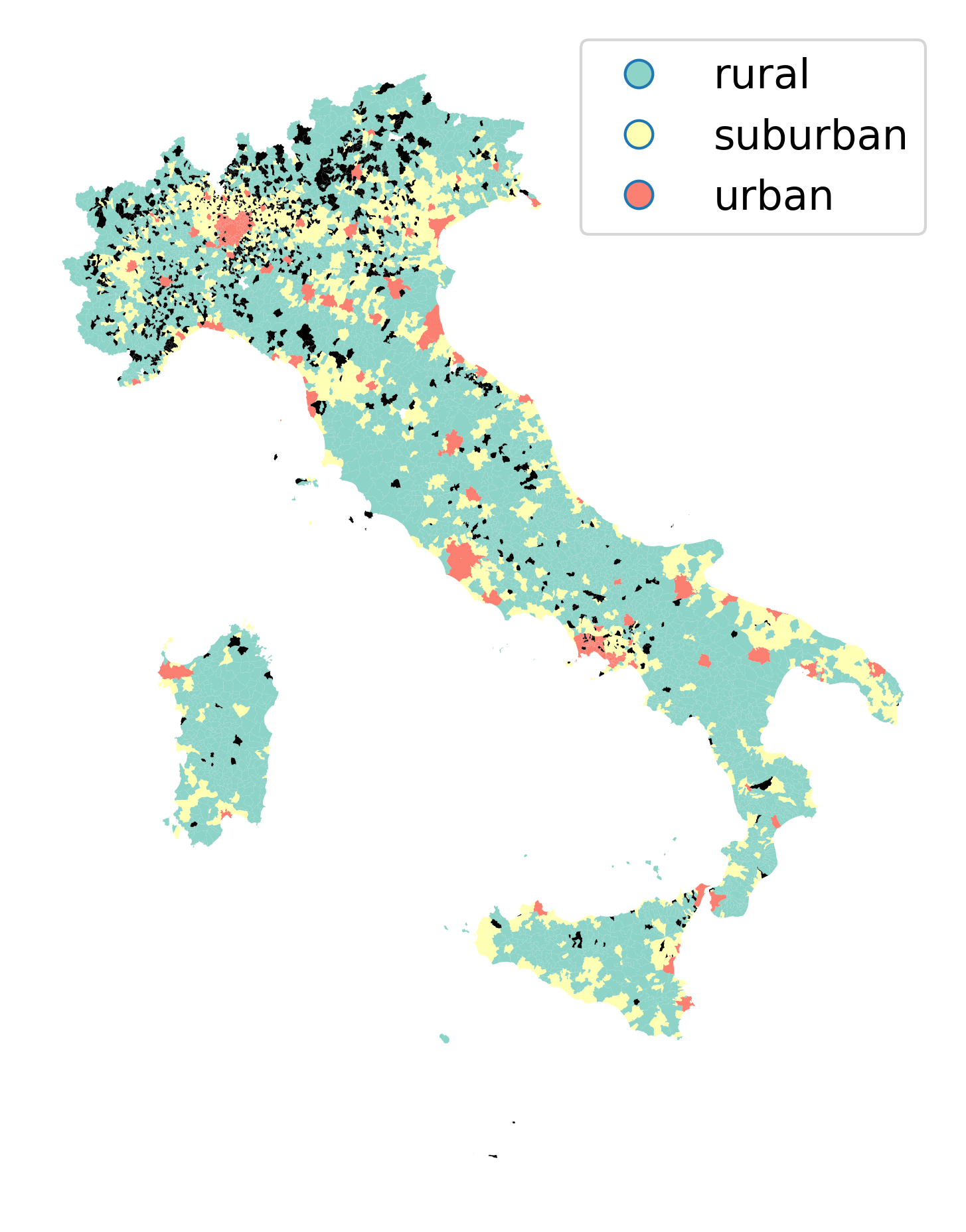}
    \caption{Italian municipalities colored according to degree of urbanization. They are colored in black if not available on Facebook marketing platform.}
    \label{fig:mapofitaly}
\end{figure}

The Italian national institute of statistics (Istat\footnote{\url{https://www.istat.it/en/}}) adopts the definition of urbanization from Eurostat, the statistical office of the European Union, and separates municipalities into three categories: cities, towns and suburbs, and rural areas\footnote{\url{https://ec.europa.eu/eurostat/web/degree-of-urbanisation/background}}. The assignment of these categories is based on population density, as measured using 1 km$^2$ grid. In Italy, there are 270 urban, 2,303 suburban, and 5,405 rural municipalities, having an average population of 74.9K, 11.2K, and 2.7K, respectively (See Figure \ref{fig:mapofitaly}). 

In order to work with the municipalities via Facebook Marketing API, we first request their IDs by specifying municipality name, region, and state and, in case more than one match is returned, use string matching to choose that with closest name. 
Out of the 7,978 Italian municipalities, we are able to match 6,891, excluding 24 (8.9\%) urban, 282 (12.2\%) suburban, 781 (14.4\%) rural municipalities. Note that matching of urban ones is easier, showing for a bias to densely populated areas even at this stage of data collection.

For all Italian municipalities, we download the demographic and socio-economic indicators from Istat. Unlike aggregate indicator values for larger geographic regions, only few are available at the fine-grained level of municipalities. Thus, we are able to collect data on the overall population (overall and split by gender), education (high school attainment and college attainment, from last census in 2011), income (net income per capita, 2018), and migration (Italian residents who are not Italian citizens, 2018). 

\subsection{Facebook Marketing API}

Facebook Advertising audience estimates are available via Facebook Marketing API, which we access using a Python package\footnote{\url{https://github.com/facebook/facebook-python-business-sdk}}.
For all queries, we request a count of ``People who live there'' (technically, setting \texttt{location\_type} parameter to \texttt{home}), as we are interested in the population living in the municipality of interest, not those working there or passing through as tourists. 
The API also allows the querying of users using other services owned by Facebook, including Instagram, Messenger, and ``Audience network''. However, we choose to constrain the query to Facebook users, for simplicity of interpretation of results. 

Furthermore, several advertising campaign types are available, focusing on either ``brand awareness'' or ``reach''. As we are interested in the most complete count of the users on the platform, we choose the ``reach'' option, which targets the ``maximum number of people''\footnote{\url{https://www.facebook.com/business/help/197976123664242}}. Finally, in the reply to our query, we save the Monthly Active Users (MAU) (a Daily Active Users count is also available, but we do not use it, as it is less stable over time).
Once we compose the queries combining the options above with various combinations of targeting options (described below), we query the Facebook Marketing API via Python, with a delay of 8 seconds empirically determined to avoid passing the rate limits.

\subsection{Targeting}

%Hypotheses and attributes/interests we use to explore them.
We build on previous literature to choose attributes for comparison of urban and rural municipalities, as well as Facebook-specific ones dealing with user interests (as inferred from user profile and activity) and technological aspects of user interactions, such as what kind of phone and connection they use to access it. We list the attributes below:

\begin{itemize}
    \item Gender (male, female)
    \item Marital status (single, married)
    \item Education (high school grad, college grad)
    \item Cell network (3G, 4G, Wi-Fi)
    \item Cellphone operating system (Android, iOS)
    \item Newness of cellphone (``Technology early adopter")
    \item Travel (living abroad, away from hometown, frequent travel, frequent international travel)
    \item Interests pertaining to culture (Catholic church, gambling)
    \item Interests pertaining to health (cooking, fast food, restaurants, fitness and wellness)
\end{itemize} % 22 total + all users

Some of these are inferred by Facebook from the self-disclosed information and from the user interactions on the platform (such as marital status and education). 
Others are determined automatically from the metadata associated with the connection, such as which cell network or phone is being used, which may be more reliable.
The resulting query consists of 6,891 municipalities, each queried once to estimate the total population, plus 22 times for populations with the above attributes. 
Note that by restricting our focus to these sub-populations, the problem of coverage becomes even more dire. In the next section we discuss our approach to solving it. An online interactive map showing the Facebook population estimates for each of the above attributes can be found at \url{http://www.datainterfaces.org/projects/facebookMap/}.

\subsection{Exclusion Query}

% Introduce exclusion queries, write about our validation of results
If the combination of targeting options for the query is too specific, the resulting MAU estimate may be limited to a lower threshold of 1,000 users. 
Given the FB variables that we chose, the standard querying process excludes 95\% rural, 67\% suburban and 38\% urban municipalities from the dataset.

To overcome this limitation, we propose an ``exclusion'' query, wherein in order to get an attribute-constrained population estimate of a small municipality \texttt{S}, it is first queried with another, larger, ``reference'' municipality \texttt{R} resulting in a combined query \texttt{S+R}. The difference between the combined query and the known reference municipality population (\texttt{(S+R)-R}) provides us an estimate for the small municipality \texttt{S} with a possible range of down to 100. This lower range is possible due to the finer resolution of results, which is not in 1,000s, but in the 100s.

Below we summarize the steps taken to query the Facebook API for municipality estimates matching a set of attributes:
\begin{enumerate}
    \item Query Facebook API for all municipalities using the standard query.
    \item Choose 5 reference municipalities with MAUs in the range between 2,000 and 10,000
    \item For each of the municipalities that previously hit the 1,000 threshold, we run combined queries 5 times, each one with a reference municipality
    \item Compute the difference of combined query (only if it did not also hit 1,000 threshold) and the reference municipality alone, and take the average across all the valid (non-negative, non-zero) responses: resulting in the ``exclusion query" estimate.
\end{enumerate}

Thus, for each collection, some queries may result in valid responses using the standard query, while others will need an exclusion query, and even these may not result in a valid estimate. However, with this approach we aim to improve the coverage of sparsely populated municipalities as well as the resolution of the estimates.

In order to assess the accuracy of the exclusion query estimates with respect to the standard ones, we select 20 municipalities for each Facebook variable and degree of urbanization for which standard estimates are known and we compared them with the same estimates extracted using the exclusion query approach. For such municipalities, the mean Pearson's correlation coefficient across all Facebook variables between the two types of estimates is 0.99, showing that exclusion query closely tracks the results of the standard one.

\section{Coverage / Stability Trade-off}

After we select the municipalities of interest and the population attributes we want to acquire from Facebook Advertising, we perform a data quality study. 
As a black box, we ask, how much does the advertisement audience estimates vary across time and populations?
Upon initial experimentation, we find the estimates change within a week of original query. 
Thus, we perform 5 data collections every two weeks in the time span between April 7th, 2019 and June 2nd 2019, and examine the variability of the results.

\subsection{Spatial coverage}

\begin{figure}[t]
   \centering
   \includegraphics[width=0.95\linewidth]{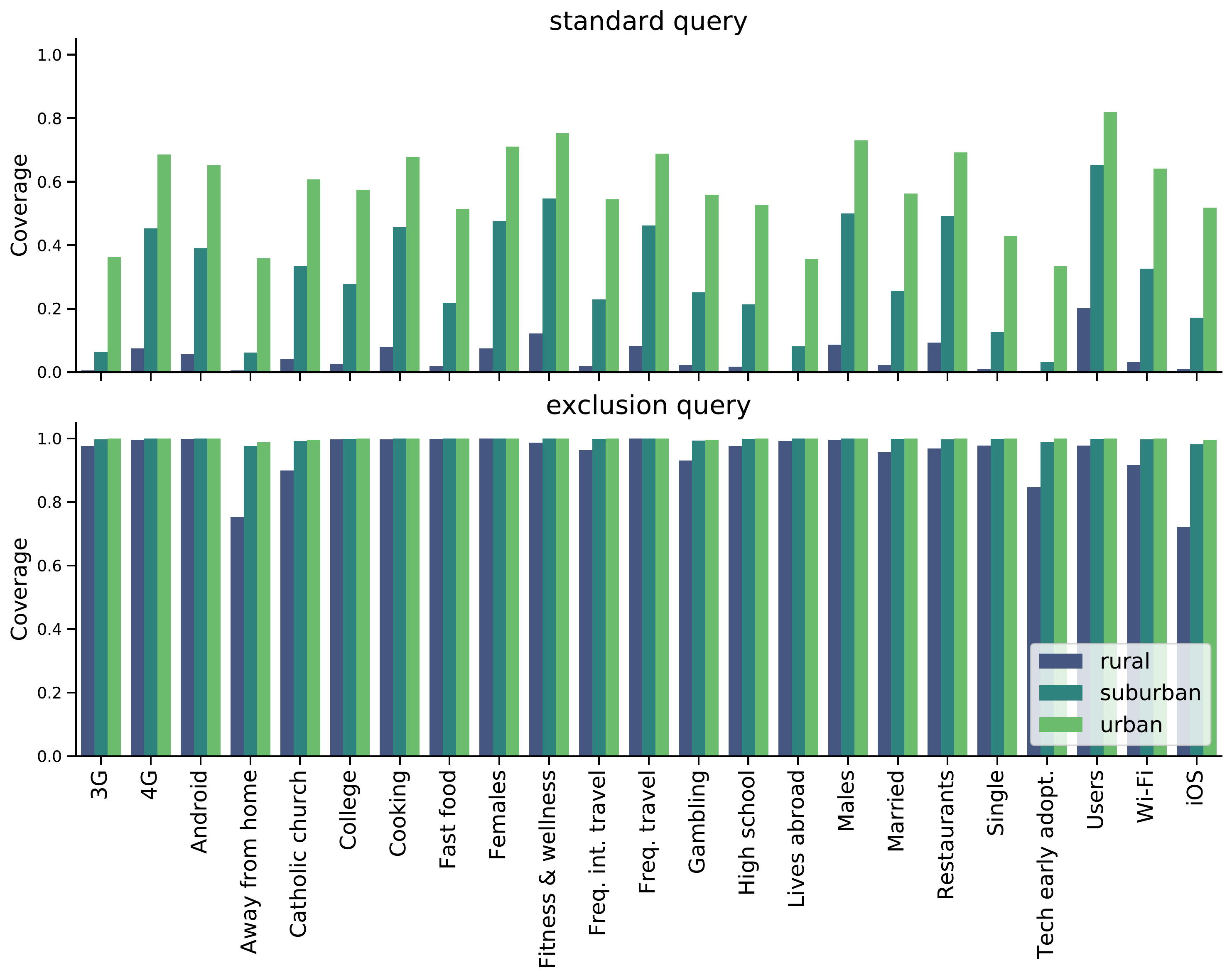}
   \caption{Coverage of municipalities reached using ``standard'' and ``exclusion'' querying with respect to the Facebook variables selected.}
    \label{fig:coverage_stand_vs_excl}
\end{figure}

\begin{figure}[t]
   \centering
   \includegraphics[width=0.95\linewidth]{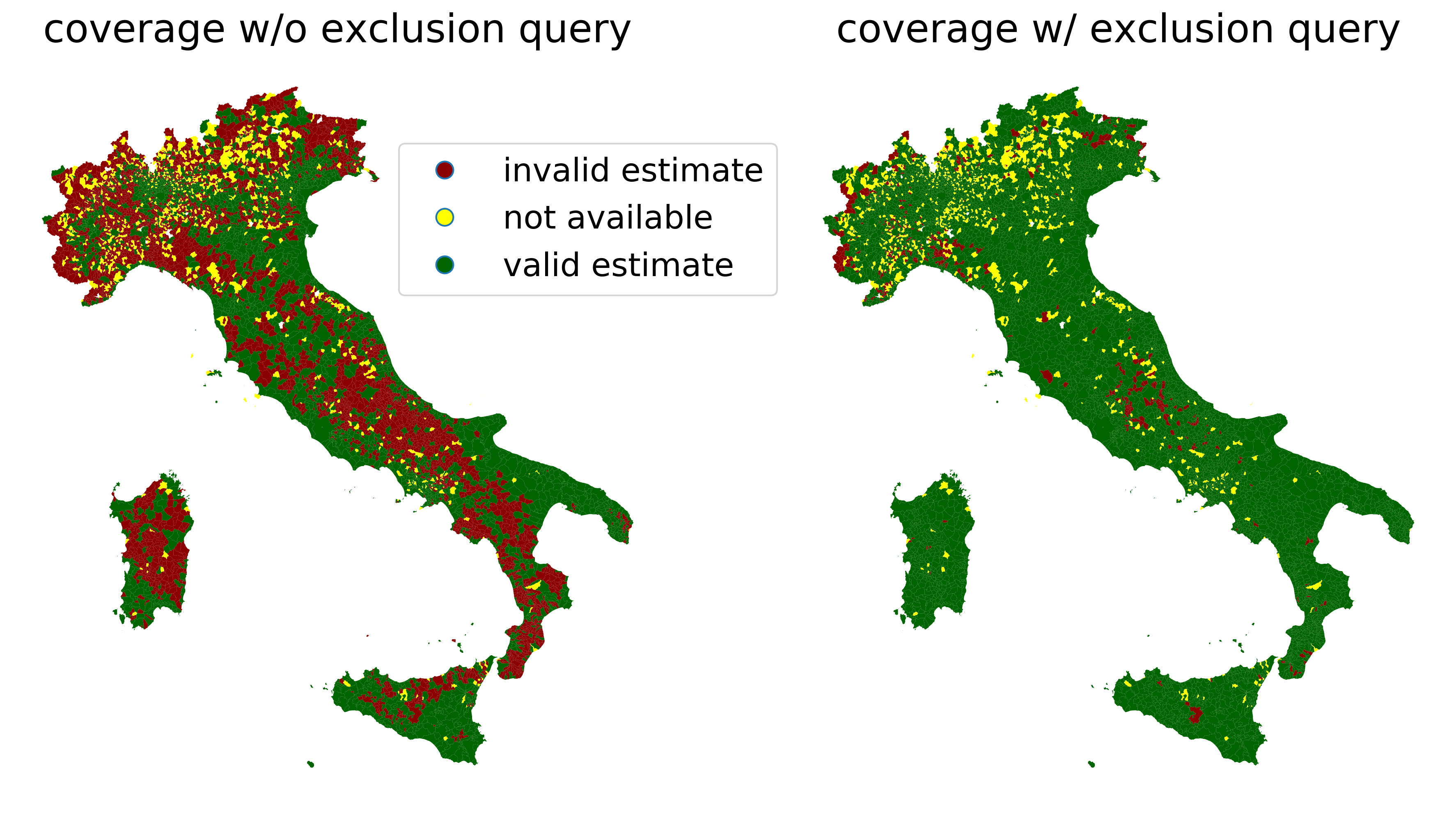}
  \caption{Coverage of municipalities within Italy, querying population without variable constraints.}
  \label{fig:coverage_users_map}
\end{figure}

First, we measure the spatial coverage for each Facebook attribute as the proportion of municipalities for which Facebook provides a valid estimate. 
We start by considering the standard query estimates, which are limited to a minimum threshold of 1,000 users. 
In this case, we consider a municipality having a valid estimate if, considering a Facebook variable, it receives at least one response above the threshold over 5 runs of the same query. 
Figure \ref{fig:coverage_stand_vs_excl} (top) shows the percentage of municipalities covered by the standard query by degree of urbanization for each Facebook variable considered. 
The variable \emph{Users} refers to the total number of Facebook users living in a municipality (without any constraints applied). 
Focusing on this variable, we observe that 81\%, 64\%, and 19\% of urban, suburban, and rural municipalities respectively are covered, showing that rural municipalities are indeed hard to reach. 
For example, in Figure \ref{fig:coverage_users_map} the map on the left shows the coverage of the standard query when querying municipality population without any attribute constraints.  
If we restrict the analysis to subgroups of Facebook users matching certain constraints, this disparity grows even more. 
While the coverage of urban municipalities is in the range between 40\% and 80\% for almost any variable, in the case of rural municipalities it is always below 15\%.

In the case of the exclusion query, the estimates now have a possible range down to 100 users. 
Therefore, to calculate the spatial coverage, we consider a municipality having a valid estimate if, given a Facebook variable, it receives at least one response of 100 or more, over 5 runs of the same query. 
Figure \ref{fig:coverage_stand_vs_excl} (bottom) illustrates the substantial improvement in the coverage for every Facebook variable selected, and \ref{fig:coverage_users_map} (right) shows the geographical coverage in the case of the generic query, suggesting that this approach can effectively be used to reach even sparsely populated municipalities.

Note that, in calculating the coverage, we considered all estimates greater or equal to 100 users as valid. Nevertheless, we may choose different cut-offs, e.g. 200, 300, 400, etc., with the corresponding change in the coverage.
To understand which threshold is most suitable, we perform a stability analysis of the results in the following section.

\subsection{Stability of query results}

\begin{figure}[t]
    \centering
    \includegraphics[width=\linewidth]{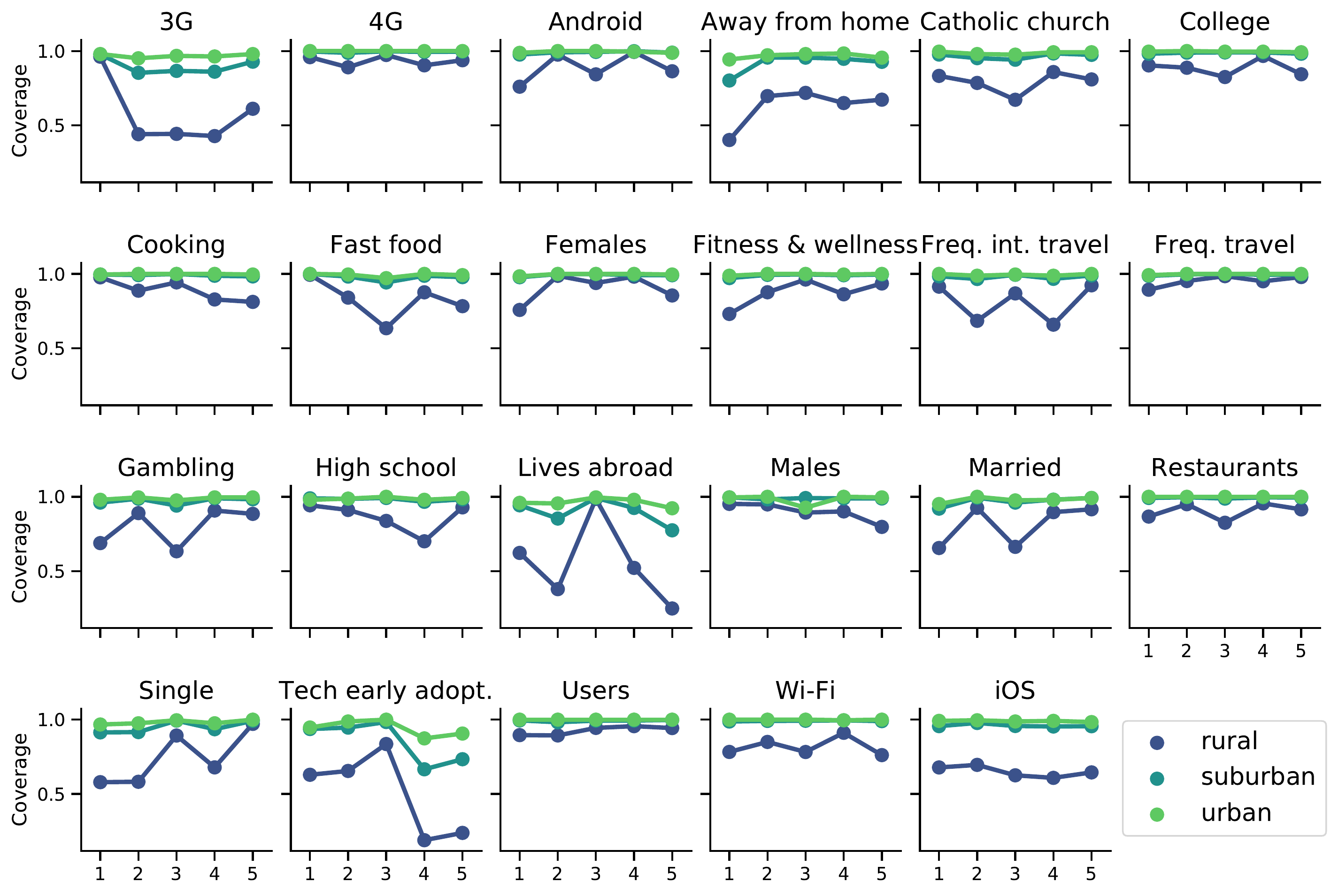}
    \caption{Percentage of municipalities having valid response for each attribute over 5 runs.}
    \label{fig:coverage_fiveruns}
\end{figure}

To assess the stability of the estimates, we examine the values of the same variables collected five times, each two weeks apart. 
First, considering a threshold of 100 users, we calculate the coverage for each collection as the percentage of all municipalities which pass the threshold. 
The result is shown in Figure \ref{fig:coverage_fiveruns}. 
While the spatial coverage is somewhat stable and close to 100\% for suburban and urban municipalities, %(except for the variable \emph{Technology early adopters} which shows a drop after the third week), 
it shows large fluctuations in the case of rural municipalities.
For instance, variable \emph{Lives abroad} ranges from almost full coverage on third day to almost no coverage for rural municipalities on the fifth.

\begin{figure}[t]
    \centering
    \includegraphics[width=\linewidth]{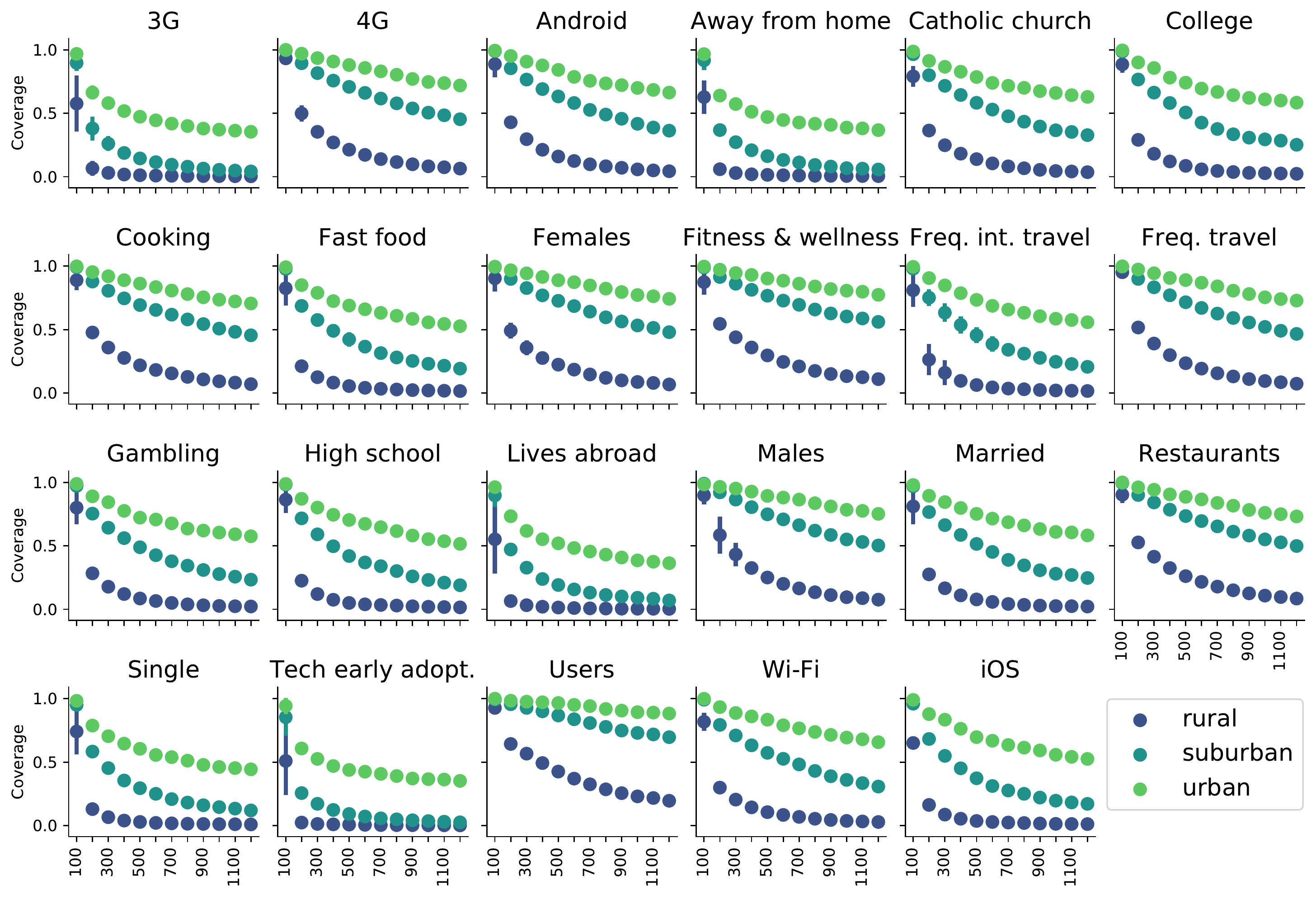}
    \caption{Coverage of municipalities for each attribute over different definitions of threshold for valid response.}
    \label{fig:coverage_thresholds}
\end{figure}

However, a threshold of 100 is the most optimistic, and we may want to consider stricter ones to improve the quality of the estimate.
We check the impact of the threshold on coverage by selecting twelve thresholds from 100 to 1,200 users with a resolution of 100 users.
Figure \ref{fig:coverage_thresholds} shows the coverage decreases smoothly for urban and suburban municipalities as the threshold increases. 
For rural (and sometimes more populated) municipalities, we observe a drop between 100 and 200 users, while after 200 users the coverage decreases smoothly, indicating the threshold of 100 may be artificially high.

\begin{figure}[t]
    \centering
    \includegraphics[width=\linewidth]{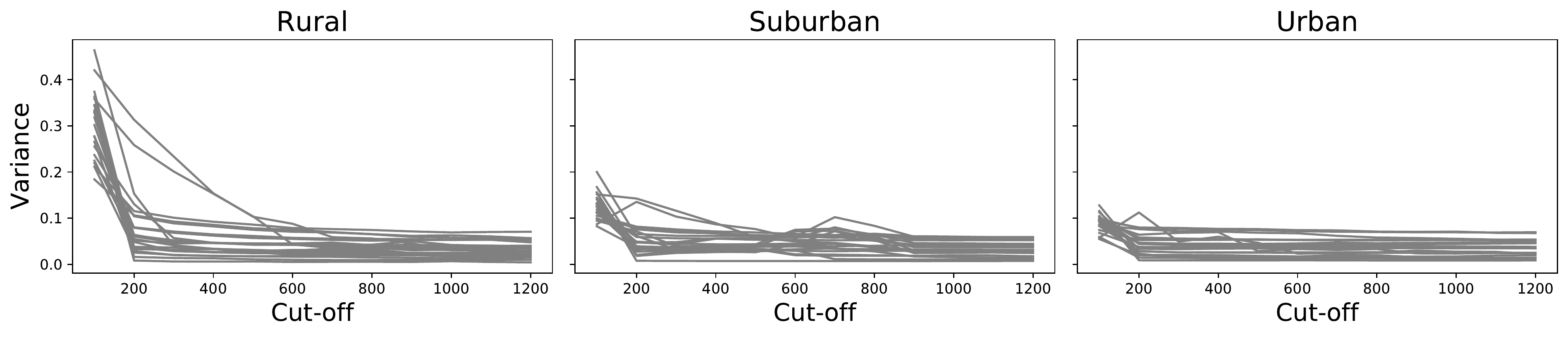}
    \caption{Variance of Facebook variables over different thresholds for valid response.}
    \label{fig:variance}
\end{figure}

Finally, we check the variability of the estimates at different thresholds.
To do this, for each Facebook variable, we compute the proportion of sub-population $P$ with specific characteristic $c$ defined as $P_m[c] = FB_m[c]/FB_m$ where $FB_m$ is the total number of users who live in municipality $m$ and $FB_m[c]$ is the total number of users who live in municipality $m$ and also match the characteristic $c$. 
Given the variability of estimates across days, $FB_m[c]$ is calculated as the median across all the valid estimates over five runs of the same query. 
Now, we calculate the variance of the distribution of this index for the three degrees of urbanization, which is shown in Figure \ref{fig:variance} (colors of the 22 attributes omitted for clarity). 
We observe that the variance rapidly decreases in the range from 100 to 200 users, while after the latter it remains stable for most of the Facebook variables.
Combined with our earlier observation of coverage drastically falling from 100 to 200 users, we choose the threshold of 200 for the following experiments in order to achieve the best coverage while ensuring the stability of the estimate.
With this threshold, we are able to cover 55\% rural, 84\% suburban, and 90\% urban municipalities, nearly tripling the coverage of the rural ones.
The improvement is even more drastic for specialized queries: the average coverage of 4\%, 29\%, and 57\% for rural, suburban, and urban municipalities now becomes 31\%, 67\%, and 81\% with the use of exclusion query.

\section{Data Quality Assessment}

\subsection{Relating to Government Statistics}

As described in Section \ref{sec:realworlddata}, some demographic and socio-economic indicators are available at the municipality level. 
To understand how well Facebook Advertising audience estimates track the figures gathered by the Italian Government (Istat), we correlate the Facebook variables most closely related to each of the Istat demographic variable in this section.

\begin{figure}
    \centering
    \includegraphics[width=0.99\linewidth]{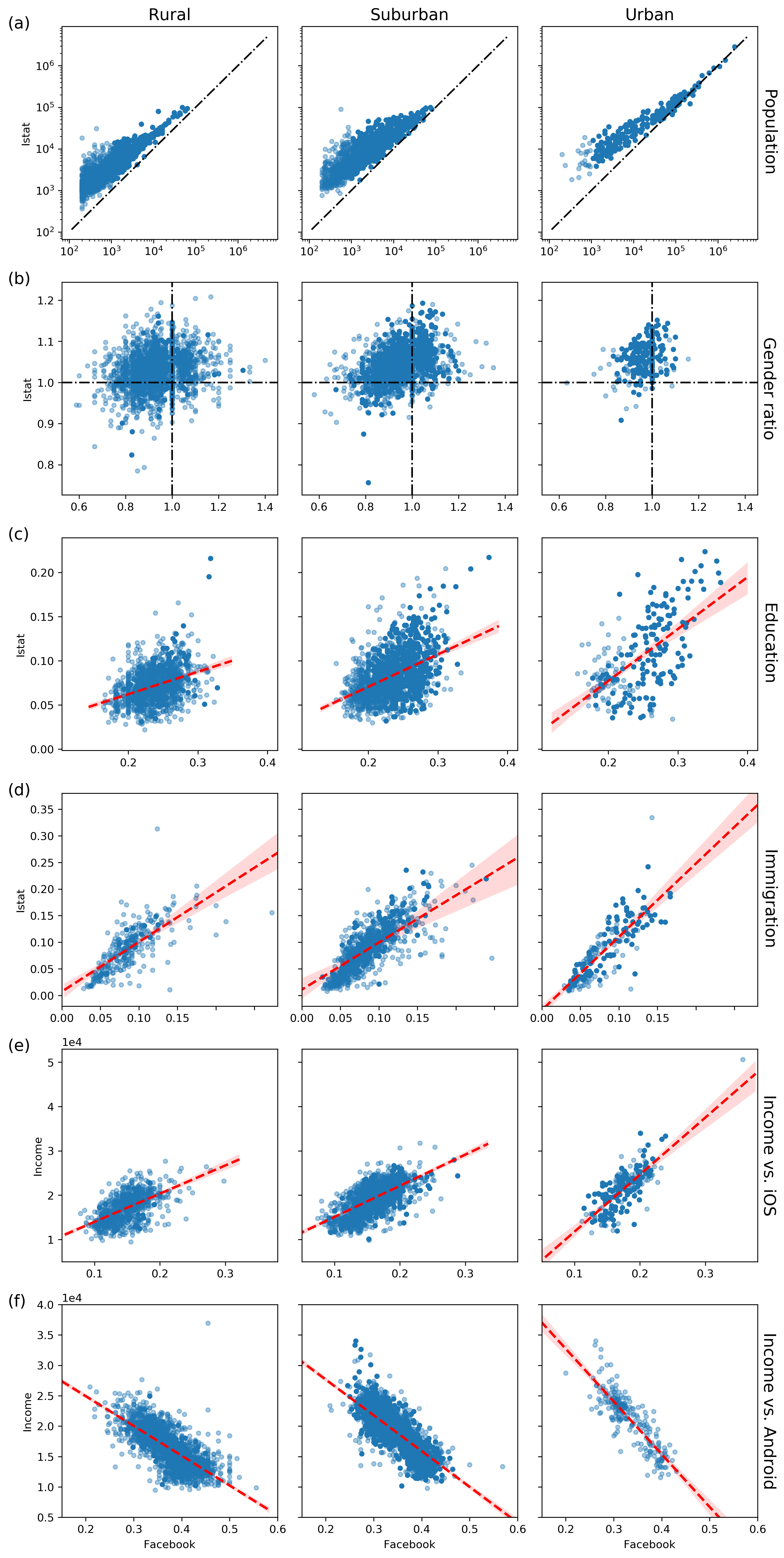}
    \caption{Facebook sub-population estimates (x axis) versus relevant Istat statistics (y axis). Dark marks are estimates of standard query, light -- exclusion query. Dashed black lines show diagonals (or parity in case of gender) and red lines show regression line with 95\% confidence.}
    \label{fig:istatplots}
\end{figure}

First set of plots in Figure \ref{fig:istatplots} shows the relationship between municipality size, as estimated using Facebook (x axis), and the population given by Istat (y axis), shown separately for rural, suburban, and urban municipalities. Note the different scale, as populations in the three plots differ. Despite high Pearson correlations of 0.93, 0.89, 0.99 for rural, suburban, and urban, respectively, we find that Facebook under-counts people in smaller municipalities, with almost all data points above the diagonal. The estimates become more accurate for the highly populated municipalities, which appear on the diagonal. The under-counting, in fact, affects the rural municipalities at a higher rate, with Facebook estimates being 71\% off on average, compared to 55\% for urban municipalities. Note that the figure distinguishes between data points acquired using standard query (in dark marks) and extended (in light marks), illustrating the drastically improved coverage of rural municipalities from 19\% to 55\% (with the same correlation). 

Considering the gender, the raw numbers are again highly correlated to the Istat population (graphs omitted for brevity, but they look much like population ones). 
Instead, we examine the gender ratios within each municipality. 
Figure \ref{fig:istatplots}b shows the comparison between the gender ratio estimated by Facebook (female/male), to that estimated by Istat, with dashed lines showing parity (50\% females and 50\% males). 
Despite actual Istat statistics showing a trend toward municipalities with more women then men, we observe that Facebook tends to overcount males, and especially so in rural communities.

Encouraged by the substantial correlation of population statistics, we examine a more challenging case of monitoring particular characteristics, measured as a proportion of population. 
For instance, Figure \ref{fig:istatplots}c shows the proportion of Facebook users who have attained a college degree, compared to the Istat estimate of the same statistic. 
The Pearson correlation between the rural, suburban, and urban areas are 0.36, 0.46, 0.61, respectively. 
Checking the outliers in the upper right of rural plot, we find Urbino and Camerino, municipalities containing universities which have a high proportion of educated residents which may be more captured by Facebook and less by formal residency requirements of Istat.
Figure \ref{fig:istatplots}d shows the proportion of Facebook users marked as ``Living abroad'', compared to the Istat estimates of ``foreigners'' residing in each municipality. 
We find substantial correlations of 0.72, 0.72, and 0.82 for rural, suburban, and urban, respectively. 
Similarly, checking outliers in the three graphs, we find that in some cases Facebook drastically over-counts the number of those ``living abroad'', and upon manual inspection a month later we find the numbers to come down, indicating a high variability either due to actual measurement of human mobility, or internal platform changes.
Finally, Figure \ref{fig:istatplots}e shows two proxies of wealth -- proportion of Facebook users who use iOS or Android devices -- compared to Istat estimates of income per capita. 
We find a clear signal that the use of iOS is positively related with income (correlations of 0.57, 0.62, 0.78) and use of Android is negatively related (-0.73, -0.77, -0.86 for rural, suburban, and urban). 
 
Note that, unlike the gender, educational attainment, and migration, the last two attributes of the phone's operating system are detected directly and unobtrusively, instead of inferred from self-reported data. 
Thus, it may not be suffering from as much measurement error as the others, and thus provide a clearer signal.

\begin{figure}[t]
    \centering
    \includegraphics[width=0.85\linewidth]{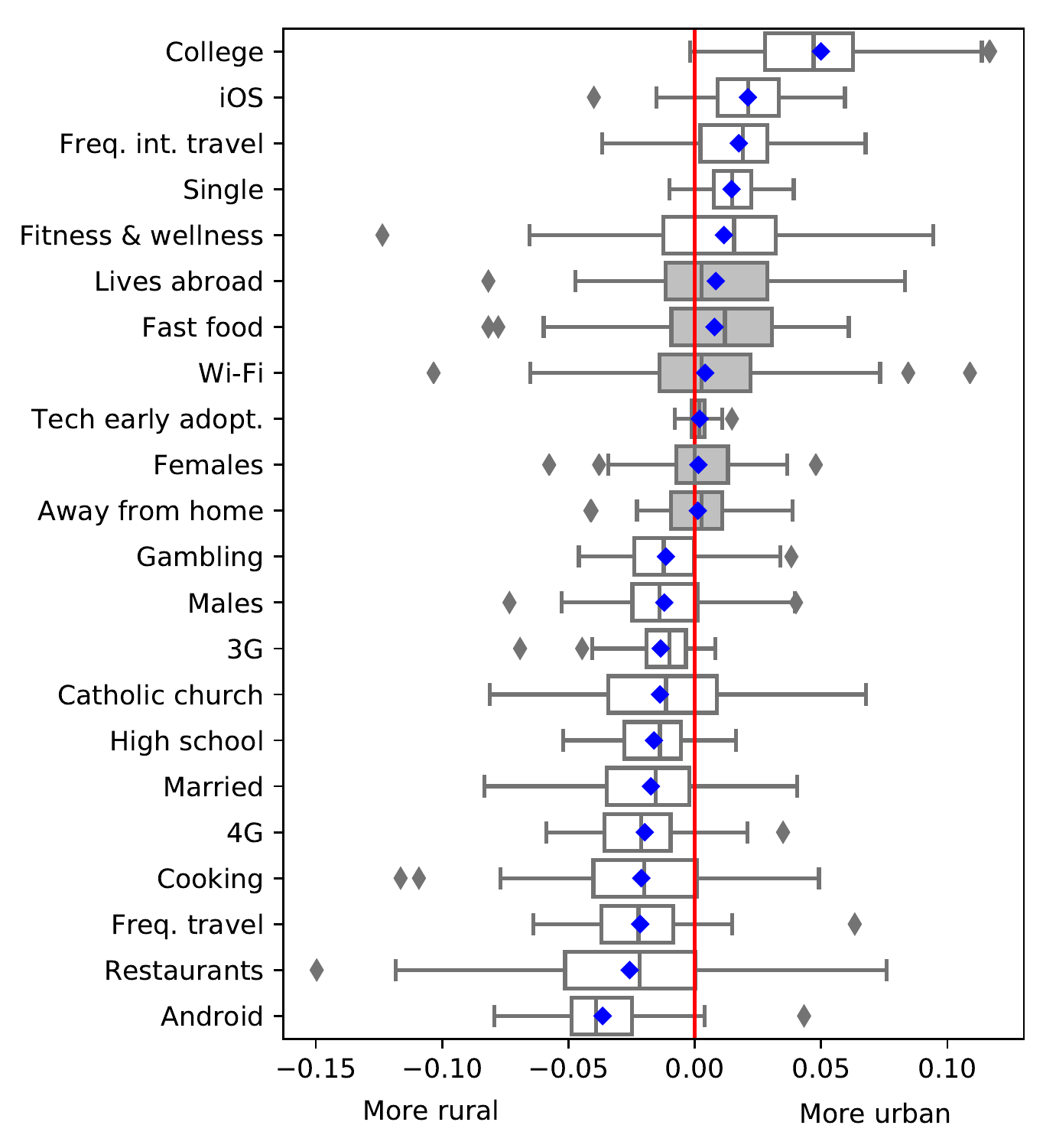}
    \caption{Distribution of difference in medians between urban and rural areas in Italian provinces for each attribute. Statistically significant boxplots are in white ($p<0.05$), insignificant in grey. Blue points show means.}
    \label{fig:inequality-provinces}
\end{figure}

\subsection{Measuring Inequality via Facebook}

Next, we use Facebook signals to measure potential inequalities between the urban and rural communities in Italy -- those that can be also tracked via Istat, and those that may be more difficult to track officially, such as interests, hobbies, travel habits, etc.

For this experiment, we exclude suburban municipalities for clarity, and leave for future work a more continuous analysis. 
For each Facebook attribute, we consider a geographically-cohesive comparison wherein in each province (there are 107 provinces in Italy, but only 71 have both rural and urban areas to compare) we subtract the median attribute value of its rural from median of its urban municipalities. 
We then plot the distribution of these differences in Figure \ref{fig:inequality-provinces}, in which the differences at the significance level of $p<0.05$ (chosen with the small number of data points in mind) are in dark grey, and the means of distributions are indicated by the blue triangle. 
Note that the significance level is affected both by the magnitude of the difference, as well as the number of provinces that have enough coverage to capture both urban and rural areas of each province, thus ranging in coverage from 27 for technology early adopters to the maximum of 68, on average 61 provinces per attribute. 
The coverage is still markedly better than if the data was used without the exclusion queries, with an average of 36, almost half as many, provinces having enough data for analysis.

The most drastic inequalities, we find, are those associated with education (there are fewer college graduates in rural areas), and income (with fewer iOS and more Android usage in rural areas). 
These findings confirm the official Istat numbers, which show significant differences in education and income. 
In a directly comparable case, Istat shows a mean difference of 7.2\%, while the difference in College graduation Facebook attribute is 5.0\% (both significant at $p<0.001$). 
Interestingly, the Facebook data does not show a significant difference between people ``Living abroad'', whereas Istat numbers show a slight difference at 4.0\%, which may be either due to Facebook usage bias or the possible ability of the platform capturing different populations, as we discuss in Discussion section.

When we examine the Facebook attributes which cannot be directly confirmed by the Istat municipality-level data, we find that in urban areas, the Facebook users tend to be more interested in fitness \& wellness, be frequent international travelers, and be single. 
In the rural areas, they tend to be interested in cooking and restaurants, to commute (be frequent travelers), to be married, be Catholic, and use 4G or 3G networks instead of Wifi (a sign of necessity for the latest efforts by Italy to extend its broadband network\footnote{\url{https://ec.europa.eu/digital-single-market/en/country-information-italy}}. 
Thus, Facebook allows us to peer into inequalities which are not captured by the governmental agencies, for instance those of relationship status, interests, and daily technology use -- all of which could be used to examine the well-being of the population holistically, in addition to the standard demographics.

\subsection{Modeling Inequality}

In the previous section we find several Facebook indicators showing inequalities between rural and urban communities, many of which are related to the socio-economic state of its residents. We ask whether the combination of these signals may be useful in modeling the financial well-being, as measured by income per capita. Not only would such a model provide an alternative, up-to-date estimate of financial well-being, it would also provide an explanatory power to gauge the possible factors associated with income inequality.

We begin by building three baseline models using the municipality-level variables made available by Istat, one for each kind of municipality (rural, suburban, and urban). Table \ref{tab:income-prediction-istat} shows the coefficients and their significance levels for the three models, as well as the number of municipalities used in the dataset (n), coverage of all possible municipalities (f), and the Adjusted R$^2$ (which corrects for the number of attributes in the model). The best performance is attained for the Urban municipalities at R$^2$=0.660, despite the smaller dataset size. The relatively poor performance of the models in rural and suburban areas (R$^2$=0.231 and R$^2$=0.281, respectively) may signify that more information is needed to differentiate between high and low income areas. Note that the performance of these small models is limited by the data available on Istat website, and may be drastically better if other variables are added. The aim of this exercise is to convey the difference in difficulty of the task between the kinds of population densities.

\begin{table}[]
\caption{Linear regression model, predicting income (in Euros) using standardized Istat variables. For each model, number of municipalities (n), coverage of all municipalities (f), and Adjusted R$^2$ are shown. Confidence levels: $p<0.001$ ***, $p<0.01$ **, $p<0.05$ *.}
\label{tab:income-prediction-istat}
{
%\footnotesize
\begin{tabular}{lrlrlrl}
\toprule
 & \multicolumn{6}{c}{Istat variables} \\
 & \multicolumn{2}{c}{Rural$_{istat}$} & \multicolumn{2}{c}{Suburban$_{istat}$} & \multicolumn{2}{c}{Urban$_{istat}$} \\
 \midrule
 & \multicolumn{2}{c}{n=5,405} & \multicolumn{2}{c}{n=2,303} & \multicolumn{2}{c}{n=270} \\
 & \multicolumn{2}{c}{f=1.00} & \multicolumn{2}{c}{f=1.00} & \multicolumn{2}{c}{f=1.00} \\
 & \multicolumn{2}{c}{R$^2$=0.231} & \multicolumn{2}{c}{R$^2$=0.281} & \multicolumn{2}{c}{R$^2$=0.660} \\
 \midrule
(Intercept) & 16,830 & *** & 19,600 & *** & 21,250 & *** \\ \Tf
males & 211 & *** & 555 & *** & 1,442 & *** \\ \Tf
high school & 1,236 & *** & 978 & *** & 2,537 & *** \\
college & 202 & *** & 914 & *** & 2,226 & *** \\ \Tf
migrants & 773 & *** & 1,009 & *** & 984 & *** \\
\bottomrule
\end{tabular}
}
\end{table}

We follow a similar setup for modeling Istat income variable using the Facebook attributes. The first three models of Table \ref{tab:income-prediction-fb} show the performance of linear regressions modeling the income (as measured by Istat) in rural, suburban, and urban municipalities, using all available attributes. The models achieve a more uniform performance with R$^2$=0.772, R$^2$=0.798, and R$^2$=0.856, respectively, with similar intercepts as the Istat models. The coefficients for the attributes which are similar to those in Istat model now reverse their sign in some cases, and lose their significance. For example, in the rural areas the model favors the information about frequent travelers, marital status, and the access to cellphone networks. In the urban areas, instead, information about college attainment, use of iOS, and interest in fitness \& wellness are more significant. Unfortunately, the increase in performance comes at a cost of coverage: only 2\% of rural and 21\% of suburban municipalities contain complete values for all features. The number of complete records is also not sufficient to perform missing value imputation, as our additional experiments reveal.

\begin{figure}[t]
    \centering
    \includegraphics[width=0.85\linewidth]{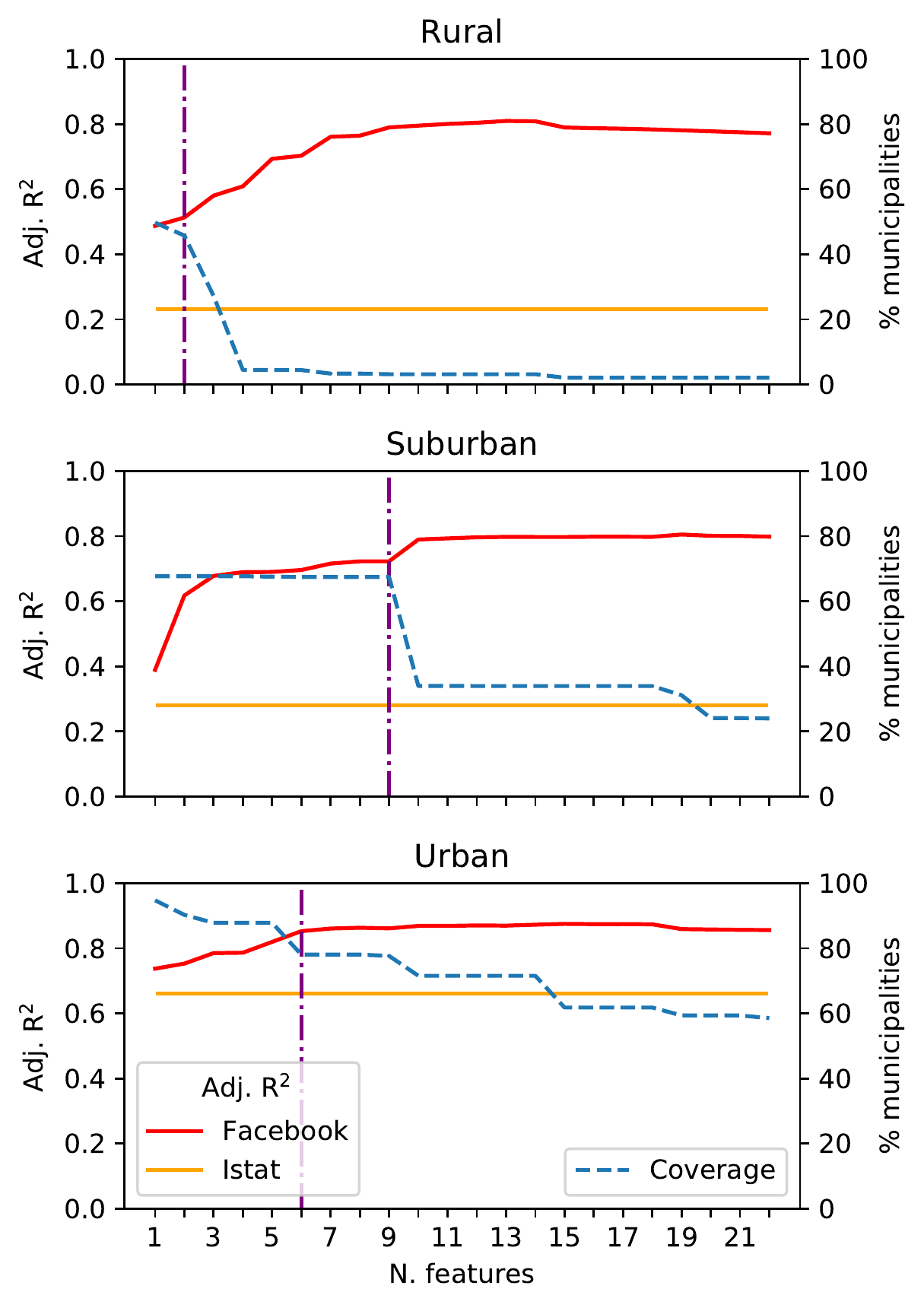}
    \caption{Coverage in \% of municipalities (right y axis) vs. performance in Adj. R$^2$ (left y axis), as features are added to the models with largest coefficients first. Baseline performance of Istat indicated by horizontal line, and the model with best trade-off indicated by vertical line.}
    \label{fig:coverage-performance}
\end{figure}

In order to improve coverage, we examine the trade-off between including features and the number of municipalities which can be used in the model. Figure \ref{fig:coverage-performance} shows the Adjusted R$^2$ performance (red line) as the number of features in the model increases (ordered by magnitude of the coefficients in the complete model), and the coverage in dashed blue line. A baseline of Istat model performance is shown as the horizontal. We observe that for rural and suburban municipalities, there comes a point when the coverage falls precipitously, with rural areas falling within the first three features. Gauging this trade-off, we select a cutoff where a reasonable performance can be achieved without discarding most of the data (shown by vertical lines). In this study we do not propose a particular metric for selecting such a cutoff, and defer the selection of such a metric to the experts in the particular issue and population being studied (where either coverage or precision may be more important).

%\todo{colors of figures are not readable in black and white}

The best trade-off models are shown in the right-most three columns of Table \ref{tab:income-prediction-fb}. The slight loss in performance is accompanied by substantial gains in coverage: from 2\% to 46\% for rural, from 21\% to 67\% in suburban, and from 53\% to 78\% in urban municipalities. In the case of urban model, the drop in performance is negligible (from R$^2$=0.856 to R$^2$=0.853). Also, while the rural model contains only two attributes, it shows substantial fitness to the data at R$^2$=0.513. Note the difference in the attributes selected for each model, showing different characteristics may be important for different levels of urbanization. 

Finally, if these models were computed on the data without using exclusion queries, there would not be enough coverage of rural municipalities to build one at all (n=11), and extremely poor coverage for suburban (n=70) and urban (n=86) municipalities.

We would caution the reader to seek a fully automated machine-learning style of optimization in this task, as the trade-off between coverage, performance, and complexity of the model (number of features) must be determined by experts in the case-by-case basis. 
Instead, we hope these experiments encourage the reader to consider sources of data alternative to the statistics gathered by governments. 

\begin{table*}[]
\caption{Linear regression models, predicting income (in Euros) using standardized Facebook variables. For each model, number of municipalities (n), coverage of all municipalities (f), and Adjusted R$^2$ are shown. Confidence levels: $p<0.001$ ***, $p<0.01$ **, $p<0.05$ *.}
\label{tab:income-prediction-fb}
{
%\footnotesize
\begin{tabular}{lrlrlrlrlrlrl}
\toprule   
 & \multicolumn{6}{c}{All Facebook attributes included}  & \multicolumn{6}{c}{Coverage vs. performance selection} \\
 \midrule
 & \multicolumn{2}{c}{Rural$_{all}$}  & \multicolumn{2}{c}{Suburban$_{all}$} & \multicolumn{2}{c}{Urban$_{all}$} & \multicolumn{2}{c}{Rural$_{cut}$}  & \multicolumn{2}{c}{Suburban$_{cut}$}  & \multicolumn{2}{c}{Urban$_{cut}$} \\
\midrule
 & \multicolumn{2}{c}{n=95} & \multicolumn{2}{c}{n=485} & \multicolumn{2}{c}{n=144} & \multicolumn{2}{c}{n=2,115} & \multicolumn{2}{c}{n=1,363} & \multicolumn{2}{c}{n=192} \\
 & \multicolumn{2}{c}{f=0.02} & \multicolumn{2}{c}{f=0.21} & \multicolumn{2}{c}{f=0.53} & \multicolumn{2}{c}{f=0.46} & \multicolumn{2}{c}{f=0.67} & \multicolumn{2}{c}{f=0.78} \\
 & \multicolumn{2}{c}{R$^2$=0.772} & \multicolumn{2}{c}{R$^2$=0.798} & \multicolumn{2}{c}{R$^2$=0.856} & \multicolumn{2}{c}{R$^2$=0.513} & \multicolumn{2}{c}{R$^2$=0.723} & \multicolumn{2}{c}{R$^2$=0.853} \\
\midrule
(Intercept) & 16,300 & *** & 18,760 & *** & 20,590 & *** & 16,380 &  & 19,200 & *** & 20,940 & \\ \Tf
males & -288 &  & -56 &  & 216 &  &  &  & -612 & *** &  & \\
married & 613 & ** & -143 & *** & 646 &  &  &  &  &  &  & \\
single & -20 &  & -159 &  & -886 & * &  &  &  &  & -1,424 & *** \\ \Tf
high school & -133 &  & 150 &  & -356 &  &  &  &  &  &  & \\
college & 265 &  & 455 & *** & 1,486 & *** &  &  & 120 &  & 2,057 & *** \\ \Tf
lives abroad & 307 &  & -11 &  & -455 &  &  &  &  &  &  & \\
away from hometown & -286 &  & 52 &  & 250 &  &  &  &  &  &  & \\
frequent international travelers & 15 &  & -281 & ** & -32 &  &  &  &  &  &  & \\
frequent travelers & -1,212 & *** & -804 & *** & -229 &  & -2,325 & *** & -841 & *** &  & \\ \Tf
android & 27 &  & -360 &  & -1,539 & * &  &  & -748 & *** & -2,011 & *** \\
ios & 384 &  & 1,140 & *** & 1,181 & ** &  &  & 982 & *** & 1,362 & *** \\
technology early adopters & 159 &  & 37 &  & -168 &  &  &  &  &  &  & \\
3g & -468 & * & -283 & ** & 14 &  &  &  &  &  &  & \\
4g & -693 & * & -607 & *** & 180 &  & -148 & * & -476 & *** &  & \\
wi-fi & 274 &  & 97 &  & -394 &  &  &  &  &  &  & \\ \Tf
retaurants & 159 &  & -71 &  & -1,094 & * &  &  &  &  & -2,122 & *** \\
fast food & -194 &  & -76 &  & 97 &  &  &  &  &  &  & \\
cooking & -261 &  & -596 & ** & -448 &  &  &  & -445 & *** &  & \\
catholic church & -114 &  & -404 & *** & -664 &  &  &  & -436 & *** &  & \\
fitness and wellness & 378 &  & 335 & ** & 1,130 & ** &  &  & 100 &  & 1,284 & *** \\
gambling & -241 &  & 25 &  & -689 &  &  &  &  &  &  & \\
\bottomrule
\end{tabular}
}
\end{table*}

% TO-DOs
% include figure with coefficients of “best model"?
% add figures of coverage vs. score

\section{Discussion \& Conclusions}

% Contribution 1 - Facebook's variables extend demography
In this work we explore an increasingly popular data source in the field of Digital Demography, the Facebook Advertising platform,
which, in addition to cost-effective population estimates, provides rich behavioral data at an unprecedented scale and granularity. 
We find that, much like the standard demographic research, it takes more effort to obtain reliable statistics for the rural populations.
Nonetheless, the rich behavioral and technical insights Facebook is able to collect on its users have a potential to extend the study of wellbeing of populations. 
For instance, health-related behaviors such as having interests in cooking at home or exercise may help in contextualizing the ongoing obesity and diabetes epidemics, much of which has been recently attributed to the rural communities \cite{bixby2019rising}. 
Further, we find the variables connected to the use of technology, and especially of mobile devices, a strong proxy to financial wellbeing of the population.
In the case of internet access, the fact that the rural residents are more likely to connect to Facebook via mobile data network instead of WiFi (land-based internet connection) may point to a persisting digital infrastructure divide between the rural and urban areas \cite{nandi2016computing}.
In aggregate, an expanded view of population's behaviors, interests, and demographics would be useful in creating compound measures of wellbeing (such as in \footnote{\url{http://lab24.ilsole24ore.com/qdv2018/}}).

% Contribution 2 - platform analysis & new query
However, our analyses uncover serious instability and coverage issues in the signal Facebook Advertising provides, especially when it pertains to the rural communities. 
The volatile behavior of this data source should be a cautionary tale to any demographers using digital platforms as ``black boxes'' that have opaque implementation and unpredictable update schedules.
The method we propose to improve the quality of data for smaller populations drastically increases the coverage of, for instance, the income model, raising the number of rural municipalities with complete records from 2\% to 46\%, allowing us to take advantage of the finer-grained attributes Facebook provides.
Though we caution the reader not to focus on the particular numbers achieved here, as they depend also on the model construction and other methodological choices, all of which should be adjusted when working on other domains and variables of interest. 
Nonetheless, a choice must be made between the coverage and signal stability, the margins of which may be best determined by the knowledge of experts in the under-represented demographic group of interest. 
When choosing the parameters, robustness checks must ensure a ``researcher degree of freedom'' \cite{wicherts2016degrees} does not lead to misleading observations and conclusions that could impact real-world policies.

% Contribution 3 - Biases and their potential sources
Besides the coverage and stability issues, we also uncover several biases in the Facebook Advertising data. 
Even when we employ the ``exclusion query'' methodology, Facebook audience estimates consistently under-count populations in rural areas and over-count males (with the gender imbalance being more pronounced in the rural municipalities). 
Many sources of bias may be at play: (1) self-selection bias in the user base of Facebook, thought to be younger and more tech savvy (but which may be shifting toward older users\footnote{\url{https://www.techspot.com/news/79082-facebook-rapidly-losing-millennials-us-user-base-down.html}}), (2) measurement bias in the sensitivity of Facebook's user attribute extraction pipeline (for example, the gender statistic may be swayed by numerous fake accounts\footnote{\url{https://www.buzzfeednews.com/article/craigsilverman/facebook-fake-accounts-afd}}), and (3) financial incentives to inflate the number of users who may see an advertisement (being an important revenue stream, Facebook's advertising revenue exceeded \$55 billion in 2018\footnote{\url{https://newsfeed.org/facebooks-revenue-exceeded-55-billion-in-2018/}}), among others already explored in literature on online data representativeness \cite{tufekci2014big,olteanu2019social,ruths2014social,sen2019total}.
Although it is not likely that Facebook will release details of its user attribute inference code, demographers may be able to adjust for the larger sample biases of Facebook user base, as recommended in \cite{cesare2018promises}.

% Contributions to other theories
Nevertheless, with the appropriate handling of certain biases of Facebook Advertising data, its benefits in terms of coverage and attribute diversity may contribute to the ongoing efforts in the theoretical understanding of the attribute variability within urbanization spectrum via the Urban Scaling Theory \cite{bettencourt2013origins}. 
Our own preliminary experiments showed scaling trends of Facebook attributes in the range of those in the existing literature \cite{bettencourt2007growth}.
Expanding the application of this theoretical framework to cultural and wellbeing aspects of populations, as measured through Facebook, would be an exciting future research direction.

% Ethical & privacy concerns
Finally, despite the aggregate nature of this data, Facebook Advertising (and many similar platforms) pose several ethical and privacy issues.
The methodology proposed in this study allows for the tracking of smaller demographic groups at a higher resolution, introducing risks especially for the more vulnerable populations and minorities, and hence should be applied with caution. 
It may be the case that Facebook needs to limit the exposure of user ``interests'' that may result in government censorship or prosecution, or targeting by other groups.
Conversely, the platform may undercount people with impairments or disabilities who are not able to use the website and who may be under-represented in its user base.
Thus, the ethics rules already established for sociological and demographic studies (such as those published by the American Sociological Association\footnote{\url{https://www.asanet.org/code-ethics}}) must be applied to those using new data sources, such as to protect the subjects of the study.

\begin{acks}
We gratefully acknowledge the support from the Lagrange Project of the ISI Foundation funded by the CRT Foundation.
\end{acks}

\FloatBarrier
\balance
\bibliographystyle{ACM-Reference-Format}
\bibliography{references}


\begin{thebibliography}{57}

%%% the Web address.

\ifx \showCODEN    \undefined \def \showCODEN     #1{\unskip}     \fi
\ifx \showDOI      \undefined \def \showDOI       #1{#1}\fi
\ifx \showISBNx    \undefined \def \showISBNx     #1{\unskip}     \fi
\ifx \showISBNxiii \undefined \def \showISBNxiii  #1{\unskip}     \fi
\ifx \showISSN     \undefined \def \showISSN      #1{\unskip}     \fi
\ifx \showLCCN     \undefined \def \showLCCN      #1{\unskip}     \fi
\ifx \shownote     \undefined \def \shownote      #1{#1}          \fi
\ifx \showarticletitle \undefined \def \showarticletitle #1{#1}   \fi
\ifx \showURL      \undefined \def \showURL       {\relax}        \fi
% invisible to TeX
\providecommand\bibfield[2]{#2}
\providecommand\bibinfo[2]{#2}
\providecommand\natexlab[1]{#1}
\providecommand\showeprint[2][]{arXiv:#2}

\bibitem[\protect\citeauthoryear{Adler, Cattuto, Kalimeri, Paolotti, Tizzoni,
  Verhulst, Yom-Tov, and Young}{Adler et~al\mbox{.}}{2019}]%
        {adler2019search}
\bibfield{author}{\bibinfo{person}{Natalia Adler}, \bibinfo{person}{Ciro
  Cattuto}, \bibinfo{person}{Kyriaki Kalimeri}, \bibinfo{person}{Daniela
  Paolotti}, \bibinfo{person}{Michele Tizzoni}, \bibinfo{person}{Stefaan
  Verhulst}, \bibinfo{person}{Elad Yom-Tov}, {and} \bibinfo{person}{Andrew
  Young}.} \bibinfo{year}{2019}\natexlab{}.
\newblock \showarticletitle{How search engine data enhance the understanding of
  determinants of suicide in {India} and inform prevention: observational
  study}.
\newblock \bibinfo{journal}{\emph{Journal of medical internet research}}
  \bibinfo{volume}{21}, \bibinfo{number}{1} (\bibinfo{year}{2019}),
  \bibinfo{pages}{e10179}.
\newblock


\bibitem[\protect\citeauthoryear{Alburez-Gutierrez, Zagheni, Aref, Gil-Clavel,
  Grow, and Negraia}{Alburez-Gutierrez et~al\mbox{.}}{2019}]%
        {alburez-gutierrez_zagheni_aref_gil-clavel_grow_negraia_2019}
\bibfield{author}{\bibinfo{person}{Diego Alburez-Gutierrez},
  \bibinfo{person}{Emilio Zagheni}, \bibinfo{person}{Samin Aref},
  \bibinfo{person}{SoÞa Gil-Clavel}, \bibinfo{person}{Andre Grow}, {and}
  \bibinfo{person}{Daniela~V Negraia}.} \bibinfo{year}{2019}\natexlab{}.
\newblock \bibinfo{title}{Demography in the Digital Era: New Data Sources for
  Population Research}.
\newblock
\newblock
\urldef\tempurl%
\url{https://doi.org/10.31235/osf.io/24jp7}
\showDOI{\tempurl}


\bibitem[\protect\citeauthoryear{Araujo, Mejova, Weber, and Benevenuto}{Araujo
  et~al\mbox{.}}{2017}]%
        {araujo2017using}
\bibfield{author}{\bibinfo{person}{Matheus Araujo}, \bibinfo{person}{Yelena
  Mejova}, \bibinfo{person}{Ingmar Weber}, {and} \bibinfo{person}{Fabricio
  Benevenuto}.} \bibinfo{year}{2017}\natexlab{}.
\newblock \showarticletitle{Using Facebook ads audiences for global lifestyle
  disease surveillance: Promises and limitations}. In
  \bibinfo{booktitle}{\emph{Proceedings of the 2017 ACM on Web Science
  Conference}}. ACM, \bibinfo{pages}{253--257}.
\newblock


\bibitem[\protect\citeauthoryear{Baranowska-Rataj, Barclay, and
  Kolk}{Baranowska-Rataj et~al\mbox{.}}{2017}]%
        {baranowska2017effect}
\bibfield{author}{\bibinfo{person}{Anna Baranowska-Rataj},
  \bibinfo{person}{Kieron Barclay}, {and} \bibinfo{person}{Martin Kolk}.}
  \bibinfo{year}{2017}\natexlab{}.
\newblock \showarticletitle{The effect of number of siblings on adult
  mortality: Evidence from Swedish registers for cohorts born between 1938 and
  1972}.
\newblock \bibinfo{journal}{\emph{Population Studies}} \bibinfo{volume}{71},
  \bibinfo{number}{1} (\bibinfo{year}{2017}), \bibinfo{pages}{43--63}.
\newblock


\bibitem[\protect\citeauthoryear{Barclay and Kolk}{Barclay and Kolk}{2017}]%
        {barclay2017long}
\bibfield{author}{\bibinfo{person}{Kieron~J Barclay} {and}
  \bibinfo{person}{Martin Kolk}.} \bibinfo{year}{2017}\natexlab{}.
\newblock \showarticletitle{The long-term cognitive and socioeconomic
  consequences of birth intervals: A within-family sibling comparison using
  Swedish register data}.
\newblock \bibinfo{journal}{\emph{Demography}} \bibinfo{volume}{54},
  \bibinfo{number}{2} (\bibinfo{year}{2017}), \bibinfo{pages}{459--484}.
\newblock


\bibitem[\protect\citeauthoryear{Beir{\'o}, Panisson, Tizzoni, and
  Cattuto}{Beir{\'o} et~al\mbox{.}}{2016}]%
        {beiro2016predicting}
\bibfield{author}{\bibinfo{person}{Mariano~G Beir{\'o}},
  \bibinfo{person}{Andr{\'e} Panisson}, \bibinfo{person}{Michele Tizzoni},
  {and} \bibinfo{person}{Ciro Cattuto}.} \bibinfo{year}{2016}\natexlab{}.
\newblock \showarticletitle{Predicting human mobility through the assimilation
  of social media traces into mobility models}.
\newblock \bibinfo{journal}{\emph{EPJ Data Science}} \bibinfo{volume}{5},
  \bibinfo{number}{1} (\bibinfo{year}{2016}), \bibinfo{pages}{30}.
\newblock


\bibitem[\protect\citeauthoryear{Bettencourt}{Bettencourt}{2013}]%
        {bettencourt2013origins}
\bibfield{author}{\bibinfo{person}{Lu{\'\i}s~MA Bettencourt}.}
  \bibinfo{year}{2013}\natexlab{}.
\newblock \showarticletitle{The origins of scaling in cities}.
\newblock \bibinfo{journal}{\emph{science}} \bibinfo{volume}{340},
  \bibinfo{number}{6139} (\bibinfo{year}{2013}), \bibinfo{pages}{1438--1441}.
\newblock


\bibitem[\protect\citeauthoryear{Bettencourt, Lobo, Helbing, K{\"u}hnert, and
  West}{Bettencourt et~al\mbox{.}}{2007}]%
        {bettencourt2007growth}
\bibfield{author}{\bibinfo{person}{Lu{\'\i}s~MA Bettencourt},
  \bibinfo{person}{Jos{\'e} Lobo}, \bibinfo{person}{Dirk Helbing},
  \bibinfo{person}{Christian K{\"u}hnert}, {and} \bibinfo{person}{Geoffrey~B
  West}.} \bibinfo{year}{2007}\natexlab{}.
\newblock \showarticletitle{Growth, innovation, scaling, and the pace of life
  in cities}.
\newblock \bibinfo{journal}{\emph{Proceedings of the national academy of
  sciences}} \bibinfo{volume}{104}, \bibinfo{number}{17}
  (\bibinfo{year}{2007}), \bibinfo{pages}{7301--7306}.
\newblock


\bibitem[\protect\citeauthoryear{Bixby, Bentham, Zhou, Di~Cesare, Paciorek,
  Collaboration, et~al\mbox{.}}{Bixby et~al\mbox{.}}{2019}]%
        {bixby2019rising}
\bibfield{author}{\bibinfo{person}{Honor Bixby}, \bibinfo{person}{James
  Bentham}, \bibinfo{person}{Bin Zhou}, \bibinfo{person}{Mariachiara
  Di~Cesare}, \bibinfo{person}{Christopher~J Paciorek}, \bibinfo{person}{NCD
  Risk~Factor Collaboration}, {et~al\mbox{.}}} \bibinfo{year}{2019}\natexlab{}.
\newblock \showarticletitle{Rising rural body-mass index is the main driver of
  the global obesity epidemic}.
\newblock \bibinfo{journal}{\emph{Nature}} \bibinfo{number}{569}
  (\bibinfo{year}{2019}), \bibinfo{pages}{260--264}.
\newblock


\bibitem[\protect\citeauthoryear{Bonanomi, Rosina, Cattuto, and
  Kalimeri}{Bonanomi et~al\mbox{.}}{2017}]%
        {bonanomi2017understanding}
\bibfield{author}{\bibinfo{person}{Andrea Bonanomi},
  \bibinfo{person}{Alessandro Rosina}, \bibinfo{person}{Ciro Cattuto}, {and}
  \bibinfo{person}{Kyriaki Kalimeri}.} \bibinfo{year}{2017}\natexlab{}.
\newblock \showarticletitle{Understanding Youth Unemployment in {Italy} via
  Social Media Data}. In \bibinfo{booktitle}{\emph{28th IUSSP international
  population conference}}.
\newblock


\bibitem[\protect\citeauthoryear{Caba{\~n}as, Cuevas, and Cuevas}{Caba{\~n}as
  et~al\mbox{.}}{2018}]%
        {cabanas2018facebook}
\bibfield{author}{\bibinfo{person}{Jos{\'e}~Gonz{\'a}lez Caba{\~n}as},
  \bibinfo{person}{{\'A}ngel Cuevas}, {and} \bibinfo{person}{Rub{\'e}n
  Cuevas}.} \bibinfo{year}{2018}\natexlab{}.
\newblock \showarticletitle{Facebook use of sensitive data for advertising in
  {Europe}}.
\newblock \bibinfo{journal}{\emph{arXiv preprint arXiv:1802.05030}}
  (\bibinfo{year}{2018}).
\newblock


\bibitem[\protect\citeauthoryear{Carter-Harris, Ellis, Warrick, and
  Rawl}{Carter-Harris et~al\mbox{.}}{2016}]%
        {carter2016beyond}
\bibfield{author}{\bibinfo{person}{Lisa Carter-Harris},
  \bibinfo{person}{Rebecca~Bartlett Ellis}, \bibinfo{person}{Adam Warrick},
  {and} \bibinfo{person}{Susan Rawl}.} \bibinfo{year}{2016}\natexlab{}.
\newblock \showarticletitle{Beyond traditional newspaper advertisement:
  leveraging Facebook-targeted advertisement to recruit long-term smokers for
  research}.
\newblock \bibinfo{journal}{\emph{Journal of medical Internet research}}
  \bibinfo{volume}{18}, \bibinfo{number}{6} (\bibinfo{year}{2016}),
  \bibinfo{pages}{e117}.
\newblock


\bibitem[\protect\citeauthoryear{Cesare, Lee, McCormick, Spiro, and
  Zagheni}{Cesare et~al\mbox{.}}{2018}]%
        {cesare2018promises}
\bibfield{author}{\bibinfo{person}{Nina Cesare}, \bibinfo{person}{Hedwig Lee},
  \bibinfo{person}{Tyler McCormick}, \bibinfo{person}{Emma Spiro}, {and}
  \bibinfo{person}{Emilio Zagheni}.} \bibinfo{year}{2018}\natexlab{}.
\newblock \showarticletitle{Promises and pitfalls of using digital traces for
  demographic research}.
\newblock \bibinfo{journal}{\emph{Demography}} \bibinfo{volume}{55},
  \bibinfo{number}{5} (\bibinfo{year}{2018}), \bibinfo{pages}{1979--1999}.
\newblock


\bibitem[\protect\citeauthoryear{Crosier, Brian, and Ben-Zeev}{Crosier
  et~al\mbox{.}}{2016}]%
        {crosier2016using}
\bibfield{author}{\bibinfo{person}{Benjamin~Sage Crosier},
  \bibinfo{person}{Rachel~Marie Brian}, {and} \bibinfo{person}{Dror Ben-Zeev}.}
  \bibinfo{year}{2016}\natexlab{}.
\newblock \showarticletitle{Using Facebook to reach people who experience
  auditory hallucinations}.
\newblock \bibinfo{journal}{\emph{Journal of medical Internet research}}
  \bibinfo{volume}{18}, \bibinfo{number}{6} (\bibinfo{year}{2016}),
  \bibinfo{pages}{e160}.
\newblock


\bibitem[\protect\citeauthoryear{De~Montjoye, Radaelli, Singh,
  et~al\mbox{.}}{De~Montjoye et~al\mbox{.}}{2015}]%
        {de2015unique}
\bibfield{author}{\bibinfo{person}{Yves-Alexandre De~Montjoye},
  \bibinfo{person}{Laura Radaelli}, \bibinfo{person}{Vivek~Kumar Singh},
  {et~al\mbox{.}}} \bibinfo{year}{2015}\natexlab{}.
\newblock \showarticletitle{Unique in the shopping mall: On the
  reidentifiability of credit card metadata}.
\newblock \bibinfo{journal}{\emph{Science}} \bibinfo{volume}{347},
  \bibinfo{number}{6221} (\bibinfo{year}{2015}), \bibinfo{pages}{536--539}.
\newblock


\bibitem[\protect\citeauthoryear{Depersin and Barthelemy}{Depersin and
  Barthelemy}{2018}]%
        {depersin2018global}
\bibfield{author}{\bibinfo{person}{Jules Depersin} {and} \bibinfo{person}{Marc
  Barthelemy}.} \bibinfo{year}{2018}\natexlab{}.
\newblock \showarticletitle{From global scaling to the dynamics of individual
  cities}.
\newblock \bibinfo{journal}{\emph{Proceedings of the National Academy of
  Sciences}} \bibinfo{volume}{115}, \bibinfo{number}{10}
  (\bibinfo{year}{2018}), \bibinfo{pages}{2317--2322}.
\newblock


\bibitem[\protect\citeauthoryear{DiMaggio, Hargittai, et~al\mbox{.}}{DiMaggio
  et~al\mbox{.}}{2001}]%
        {dimaggio2001digital}
\bibfield{author}{\bibinfo{person}{Paul DiMaggio}, \bibinfo{person}{Eszter
  Hargittai}, {et~al\mbox{.}}} \bibinfo{year}{2001}\natexlab{}.
\newblock \showarticletitle{From the {\textquoteleft}digital
  divide{\textquoteright}to {\textquoteleft}digital
  inequality{\textquoteright}: Studying Internet use as penetration increases}.
\newblock \bibinfo{journal}{\emph{Princeton: Center for Arts and Cultural
  Policy Studies, Woodrow Wilson School, Princeton University}}
  \bibinfo{volume}{4}, \bibinfo{number}{1} (\bibinfo{year}{2001}),
  \bibinfo{pages}{4--2}.
\newblock


\bibitem[\protect\citeauthoryear{Dubois, Zagheni, Garimella, and Weber}{Dubois
  et~al\mbox{.}}{2018}]%
        {dubois2018studying}
\bibfield{author}{\bibinfo{person}{Antoine Dubois}, \bibinfo{person}{Emilio
  Zagheni}, \bibinfo{person}{Kiran Garimella}, {and} \bibinfo{person}{Ingmar
  Weber}.} \bibinfo{year}{2018}\natexlab{}.
\newblock \showarticletitle{Studying migrant assimilation through facebook
  interests}. In \bibinfo{booktitle}{\emph{International Conference on Social
  Informatics}}. Springer, \bibinfo{pages}{51--60}.
\newblock


\bibitem[\protect\citeauthoryear{Fatehkia, Kashyap, and Weber}{Fatehkia
  et~al\mbox{.}}{2018}]%
        {fatehkia2018using}
\bibfield{author}{\bibinfo{person}{Masoomali Fatehkia}, \bibinfo{person}{Ridhi
  Kashyap}, {and} \bibinfo{person}{Ingmar Weber}.}
  \bibinfo{year}{2018}\natexlab{}.
\newblock \showarticletitle{Using Facebook ad data to track the global digital
  gender gap}.
\newblock \bibinfo{journal}{\emph{World Development}}  \bibinfo{volume}{107}
  (\bibinfo{year}{2018}), \bibinfo{pages}{189--209}.
\newblock


\bibitem[\protect\citeauthoryear{Fatehkia, O{\textquoteright}Brien, and
  Weber}{Fatehkia et~al\mbox{.}}{2019}]%
        {fatehkia2019correlated}
\bibfield{author}{\bibinfo{person}{Masoomali Fatehkia}, \bibinfo{person}{Dan
  O{\textquoteright}Brien}, {and} \bibinfo{person}{Ingmar Weber}.}
  \bibinfo{year}{2019}\natexlab{}.
\newblock \showarticletitle{Correlated impulses: Using Facebook interests to
  improve predictions of crime rates in urban areas}.
\newblock \bibinfo{journal}{\emph{PLOS ONE}} \bibinfo{volume}{14},
  \bibinfo{number}{2} (\bibinfo{year}{2019}), \bibinfo{pages}{1--16}.
\newblock
\urldef\tempurl%
\url{https://doi.org/10.1371/journal.pone.0211350}
\showDOI{\tempurl}


\bibitem[\protect\citeauthoryear{Garcia, Kassa, Cuevas, Cebrian, Moro, Rahwan,
  and Cuevas}{Garcia et~al\mbox{.}}{2018}]%
        {garcia2018analyzing}
\bibfield{author}{\bibinfo{person}{David Garcia}, \bibinfo{person}{Yonas~Mitike
  Kassa}, \bibinfo{person}{Angel Cuevas}, \bibinfo{person}{Manuel Cebrian},
  \bibinfo{person}{Esteban Moro}, \bibinfo{person}{Iyad Rahwan}, {and}
  \bibinfo{person}{Ruben Cuevas}.} \bibinfo{year}{2018}\natexlab{}.
\newblock \showarticletitle{Analyzing gender inequality through large-scale
  Facebook advertising data}.
\newblock \bibinfo{journal}{\emph{Proceedings of the National Academy of
  Sciences}} \bibinfo{volume}{115}, \bibinfo{number}{27}
  (\bibinfo{year}{2018}), \bibinfo{pages}{6958--6963}.
\newblock


\bibitem[\protect\citeauthoryear{Gil-Clavel and Zagheni}{Gil-Clavel and
  Zagheni}{2019}]%
        {gilclavel2019demographic}
\bibfield{author}{\bibinfo{person}{Sofia Gil-Clavel} {and}
  \bibinfo{person}{Emilio Zagheni}.} \bibinfo{year}{2019}\natexlab{}.
\newblock \showarticletitle{Demographic Differentials in Facebook Usage Around
  the World}. In \bibinfo{booktitle}{\emph{Proceedings of the International
  AAAI Conference on Web and Social Media}}, Vol.~\bibinfo{volume}{13(01)}.
  AAAI, \bibinfo{pages}{647--650}.
\newblock


\bibitem[\protect\citeauthoryear{Kalimeri, Beir{\'o}, Delfino, Raleigh, and
  Cattuto}{Kalimeri et~al\mbox{.}}{2019a}]%
        {kalimeri2019predicting}
\bibfield{author}{\bibinfo{person}{Kyriaki Kalimeri},
  \bibinfo{person}{Mariano~G Beir{\'o}}, \bibinfo{person}{Matteo Delfino},
  \bibinfo{person}{Robert Raleigh}, {and} \bibinfo{person}{Ciro Cattuto}.}
  \bibinfo{year}{2019}\natexlab{a}.
\newblock \showarticletitle{Predicting demographics, moral foundations, and
  human values from digital behaviours}.
\newblock \bibinfo{journal}{\emph{Computers in Human Behavior}}
  \bibinfo{volume}{92} (\bibinfo{year}{2019}), \bibinfo{pages}{428--445}.
\newblock


\bibitem[\protect\citeauthoryear{Kalimeri, G~Beir{\'o}, Urbinati, Bonanomi,
  Rosina, and Cattuto}{Kalimeri et~al\mbox{.}}{2019b}]%
        {kalimeri2019human}
\bibfield{author}{\bibinfo{person}{Kyriaki Kalimeri}, \bibinfo{person}{Mariano
  G~Beir{\'o}}, \bibinfo{person}{Alessandra Urbinati}, \bibinfo{person}{Andrea
  Bonanomi}, \bibinfo{person}{Alessandro Rosina}, {and} \bibinfo{person}{Ciro
  Cattuto}.} \bibinfo{year}{2019}\natexlab{b}.
\newblock \showarticletitle{Human Values and Attitudes towards Vaccination in
  Social Media}. In \bibinfo{booktitle}{\emph{Companion Proceedings of The 2019
  World Wide Web Conference}}. ACM, \bibinfo{pages}{248--254}.
\newblock


\bibitem[\protect\citeauthoryear{Kent and Capello~Jr}{Kent and
  Capello~Jr}{2013}]%
        {kent2013spatial}
\bibfield{author}{\bibinfo{person}{Joshua~D Kent} {and}
  \bibinfo{person}{Henry~T Capello~Jr}.} \bibinfo{year}{2013}\natexlab{}.
\newblock \showarticletitle{Spatial patterns and demographic indicators of
  effective social media content during theHorsethief Canyon fire of 2012}.
\newblock \bibinfo{journal}{\emph{Cartography and Geographic Information
  Science}} \bibinfo{volume}{40}, \bibinfo{number}{2} (\bibinfo{year}{2013}),
  \bibinfo{pages}{78--89}.
\newblock


\bibitem[\protect\citeauthoryear{Keuschnigg, Mutgan, and
  Hedstr{\"o}m}{Keuschnigg et~al\mbox{.}}{2019}]%
        {keuschnigg2019urban}
\bibfield{author}{\bibinfo{person}{Marc Keuschnigg}, \bibinfo{person}{Selcan
  Mutgan}, {and} \bibinfo{person}{Peter Hedstr{\"o}m}.}
  \bibinfo{year}{2019}\natexlab{}.
\newblock \showarticletitle{Urban scaling and the regional divide}.
\newblock \bibinfo{journal}{\emph{Science advances}} \bibinfo{volume}{5},
  \bibinfo{number}{1} (\bibinfo{year}{2019}), \bibinfo{pages}{eaav0042}.
\newblock


\bibitem[\protect\citeauthoryear{Larsen}{Larsen}{2011}]%
        {larsen2011welcome}
\bibfield{author}{\bibinfo{person}{Steph Larsen}.}
  \bibinfo{year}{2011}\natexlab{}.
\newblock \bibinfo{title}{{Welcome to the food deserts of rural America}}.
\newblock
\newblock
\newblock
\shownote{https://grist.org/article/2011-01-21-welcome-to-the-food-deserts-of-rural-america/.}


\bibitem[\protect\citeauthoryear{Mainous and Kohrs}{Mainous and Kohrs}{1995}]%
        {mainous1995comparison}
\bibfield{author}{\bibinfo{person}{Arch~G Mainous} {and}
  \bibinfo{person}{Francis~P Kohrs}.} \bibinfo{year}{1995}\natexlab{}.
\newblock \showarticletitle{A comparison of health status between rural and
  urban adults}.
\newblock \bibinfo{journal}{\emph{Journal of Community Health}}
  \bibinfo{volume}{20}, \bibinfo{number}{5} (\bibinfo{year}{1995}),
  \bibinfo{pages}{423--431}.
\newblock


\bibitem[\protect\citeauthoryear{Mejova, Gandhi, Rafaliya, Sitapara, Kashyap,
  and Weber}{Mejova et~al\mbox{.}}{2018a}]%
        {mejova2018measuring}
\bibfield{author}{\bibinfo{person}{Yelena Mejova}, \bibinfo{person}{Harsh~Rajiv
  Gandhi}, \bibinfo{person}{Tejas~Jivanbhai Rafaliya},
  \bibinfo{person}{Mayank~Rameshbhai Sitapara}, \bibinfo{person}{Ridhi
  Kashyap}, {and} \bibinfo{person}{Ingmar Weber}.}
  \bibinfo{year}{2018}\natexlab{a}.
\newblock \showarticletitle{Measuring Subnational Digital Gender Inequality in
  {India} through Gender Gaps in Facebook Use}. In
  \bibinfo{booktitle}{\emph{Proceedings of the 1\textsuperscript{st} ACM SIGCAS
  Conference on Computing and Sustainable Societies}}. ACM,
  \bibinfo{pages}{43}.
\newblock


\bibitem[\protect\citeauthoryear{Mejova and Kalimeri}{Mejova and
  Kalimeri}{2019}]%
        {mejova2019effect}
\bibfield{author}{\bibinfo{person}{Yelena Mejova} {and}
  \bibinfo{person}{Kyriaki Kalimeri}.} \bibinfo{year}{2019}\natexlab{}.
\newblock \showarticletitle{Effect of Values and Technology Use on Exercise:
  Implications for Personalized Behavior Change Interventions}. In
  \bibinfo{booktitle}{\emph{Proceedings of the 27\textsuperscript{th} ACM
  Conference on User Modeling, Adaptation and Personalization}}. ACM,
  \bibinfo{pages}{36--45}.
\newblock


\bibitem[\protect\citeauthoryear{Mejova, Weber, and Fernandez-Luque}{Mejova
  et~al\mbox{.}}{2018b}]%
        {mejova2018online}
\bibfield{author}{\bibinfo{person}{Yelena Mejova}, \bibinfo{person}{Ingmar
  Weber}, {and} \bibinfo{person}{Luis Fernandez-Luque}.}
  \bibinfo{year}{2018}\natexlab{b}.
\newblock \showarticletitle{Online health monitoring using Facebook
  advertisement audience estimates in the {United States}: Evaluation study}.
\newblock \bibinfo{journal}{\emph{JMIR public health and surveillance}}
  \bibinfo{volume}{4}, \bibinfo{number}{1} (\bibinfo{year}{2018}),
  \bibinfo{pages}{e30}.
\newblock


\bibitem[\protect\citeauthoryear{Murdock and Swanson}{Murdock and
  Swanson}{2008}]%
        {murdock2008applied}
\bibfield{author}{\bibinfo{person}{S Murdock} {and} \bibinfo{person}{DA
  Swanson}.} \bibinfo{year}{2008}\natexlab{}.
\newblock \bibinfo{title}{Applied demography in the twenty-first century}.
\newblock
\newblock


\bibitem[\protect\citeauthoryear{Murdock, Cline, and Zey}{Murdock
  et~al\mbox{.}}{2012}]%
        {Murdock2012}
\bibfield{author}{\bibinfo{person}{Steve~H. Murdock}, \bibinfo{person}{Michael
  Cline}, {and} \bibinfo{person}{Mary Zey}.} \bibinfo{year}{2012}\natexlab{}.
\newblock \bibinfo{booktitle}{\emph{Challenges in the Analysis of Rural
  Populations in the {United States}}}.
\newblock \bibinfo{publisher}{Springer Netherlands},
  \bibinfo{address}{Dordrecht}, \bibinfo{pages}{7--15}.
\newblock
\showISBNx{978-94-007-1842-5}
\urldef\tempurl%
\url{https://doi.org/10.1007/978-94-007-1842-5_2}
\showDOI{\tempurl}


\bibitem[\protect\citeauthoryear{Nandi, Thota, Nag, Divyasukhananda, Goswami,
  Aravindakshan, Rodriguez, and Mukherjee}{Nandi et~al\mbox{.}}{2016}]%
        {nandi2016computing}
\bibfield{author}{\bibinfo{person}{Somen Nandi}, \bibinfo{person}{Saigopal
  Thota}, \bibinfo{person}{Avishek Nag}, \bibinfo{person}{Sw Divyasukhananda},
  \bibinfo{person}{Partha Goswami}, \bibinfo{person}{Ashwin Aravindakshan},
  \bibinfo{person}{Raymond Rodriguez}, {and} \bibinfo{person}{Biswanath
  Mukherjee}.} \bibinfo{year}{2016}\natexlab{}.
\newblock \showarticletitle{Computing for rural empowerment: enabled by
  last-mile telecommunications}.
\newblock \bibinfo{journal}{\emph{IEEE Communications Magazine}}
  \bibinfo{volume}{54}, \bibinfo{number}{6} (\bibinfo{year}{2016}),
  \bibinfo{pages}{102--109}.
\newblock


\bibitem[\protect\citeauthoryear{OECD}{OECD}{2018}]%
        {OECD2018migrant}
\bibfield{author}{\bibinfo{person}{OECD}.} \bibinfo{year}{2018}\natexlab{}.
\newblock \bibinfo{booktitle}{\emph{OECD Regions and Cities at a Glance 2018}}.
\newblock 168 pages.
\newblock
\urldef\tempurl%
\url{https://doi.org/10.1787/reg_cit_glance-2018-en}
\showDOI{\tempurl}


\bibitem[\protect\citeauthoryear{{OECD}}{{OECD}}{2018}]%
        {OECD2018}
\bibfield{author}{\bibinfo{person}{{OECD}}.} \bibinfo{year}{2018}\natexlab{}.
\newblock \bibinfo{booktitle}{\emph{{Rural 3.0~{A} framework for rural
  develppment}}}.
\newblock \bibinfo{type}{{T}echnical {R}eport}. \bibinfo{institution}{OECD}.
\newblock
\urldef\tempurl%
\url{https://www.oecd.org/cfe/regional-policy/Rural-3.0-Policy-Note.pdf}
\showURL{%
\tempurl}


\bibitem[\protect\citeauthoryear{Olteanu, Castillo, Diaz, and Kiciman}{Olteanu
  et~al\mbox{.}}{2019}]%
        {olteanu2019social}
\bibfield{author}{\bibinfo{person}{Alexandra Olteanu}, \bibinfo{person}{Carlos
  Castillo}, \bibinfo{person}{Fernando Diaz}, {and} \bibinfo{person}{Emre
  Kiciman}.} \bibinfo{year}{2019}\natexlab{}.
\newblock \showarticletitle{Social data: Biases, methodological pitfalls, and
  ethical boundaries}.
\newblock \bibinfo{journal}{\emph{Frontiers in Big Data}}  \bibinfo{volume}{2}
  (\bibinfo{year}{2019}), \bibinfo{pages}{13}.
\newblock


\bibitem[\protect\citeauthoryear{P{\"o}tzschke and Braun}{P{\"o}tzschke and
  Braun}{2017}]%
        {potzschke2017migrant}
\bibfield{author}{\bibinfo{person}{Steffen P{\"o}tzschke} {and}
  \bibinfo{person}{Michael Braun}.} \bibinfo{year}{2017}\natexlab{}.
\newblock \showarticletitle{Migrant sampling using facebook advertisements: A
  case study of polish migrants in four {E}uropean countries}.
\newblock \bibinfo{journal}{\emph{Social Science Computer Review}}
  \bibinfo{volume}{35}, \bibinfo{number}{5} (\bibinfo{year}{2017}),
  \bibinfo{pages}{633--653}.
\newblock


\bibitem[\protect\citeauthoryear{Rampazzo, Zagheni, Weber, Testa, and
  Billari}{Rampazzo et~al\mbox{.}}{2018}]%
        {rampazzo2018mater}
\bibfield{author}{\bibinfo{person}{Francesco Rampazzo}, \bibinfo{person}{Emilio
  Zagheni}, \bibinfo{person}{Ingmar Weber}, \bibinfo{person}{Maria~Rita Testa},
  {and} \bibinfo{person}{Francesco Billari}.} \bibinfo{year}{2018}\natexlab{}.
\newblock \showarticletitle{Mater certa est, pater numquam: What can Facebook
  Advertising Data Tell Us about Male Fertility Rates?}. In
  \bibinfo{booktitle}{\emph{Twelfth International AAAI Conference on Web and
  Social Media}}.
\newblock


\bibitem[\protect\citeauthoryear{Rigg, Bebbington, Gough, Bryceson, Agergaard,
  Fold, and Tacoli}{Rigg et~al\mbox{.}}{2009}]%
        {rigg2009world}
\bibfield{author}{\bibinfo{person}{Jonathan Rigg}, \bibinfo{person}{Anthony
  Bebbington}, \bibinfo{person}{Katherine~V Gough}, \bibinfo{person}{Deborah~F
  Bryceson}, \bibinfo{person}{Jytte Agergaard}, \bibinfo{person}{Niels Fold},
  {and} \bibinfo{person}{Cecilia Tacoli}.} \bibinfo{year}{2009}\natexlab{}.
\newblock \showarticletitle{The World Development Report 2009'reshapes economic
  geography': geographical reflections}.
\newblock \bibinfo{journal}{\emph{Transactions of the Institute of British
  Geographers}} \bibinfo{volume}{34}, \bibinfo{number}{2}
  (\bibinfo{year}{2009}), \bibinfo{pages}{128--136}.
\newblock


\bibitem[\protect\citeauthoryear{Roessger, Greenleaf, and Hoggan}{Roessger
  et~al\mbox{.}}{2017}]%
        {roessger2017using}
\bibfield{author}{\bibinfo{person}{Kevin~M Roessger}, \bibinfo{person}{Arie
  Greenleaf}, {and} \bibinfo{person}{Chad Hoggan}.}
  \bibinfo{year}{2017}\natexlab{}.
\newblock \showarticletitle{Using data collection apps and single-case designs
  to research transformative learning in adults}.
\newblock \bibinfo{journal}{\emph{Journal of Adult and Continuing Education}}
  \bibinfo{volume}{23}, \bibinfo{number}{2} (\bibinfo{year}{2017}),
  \bibinfo{pages}{206--225}.
\newblock


\bibitem[\protect\citeauthoryear{Ruths and Pfeffer}{Ruths and Pfeffer}{2014}]%
        {ruths2014social}
\bibfield{author}{\bibinfo{person}{Derek Ruths} {and}
  \bibinfo{person}{J{\"u}rgen Pfeffer}.} \bibinfo{year}{2014}\natexlab{}.
\newblock \showarticletitle{Social media for large studies of behavior}.
\newblock \bibinfo{journal}{\emph{Science}} \bibinfo{volume}{346},
  \bibinfo{number}{6213} (\bibinfo{year}{2014}), \bibinfo{pages}{1063--1064}.
\newblock


\bibitem[\protect\citeauthoryear{Sahn and Stifel}{Sahn and Stifel}{2003}]%
        {sahn2003urban}
\bibfield{author}{\bibinfo{person}{David~E Sahn} {and} \bibinfo{person}{David~C
  Stifel}.} \bibinfo{year}{2003}\natexlab{}.
\newblock \showarticletitle{Urban--rural inequality in living standards in
  {Africa}}.
\newblock \bibinfo{journal}{\emph{Journal of African Economies}}
  \bibinfo{volume}{12}, \bibinfo{number}{4} (\bibinfo{year}{2003}),
  \bibinfo{pages}{564--597}.
\newblock


\bibitem[\protect\citeauthoryear{Saikia, Singh, Jasilionis, and
  Faujdar~Ram}{Saikia et~al\mbox{.}}{2013}]%
        {saikia2013explaining}
\bibfield{author}{\bibinfo{person}{Nandita Saikia}, \bibinfo{person}{Abhishek
  Singh}, \bibinfo{person}{Domantas Jasilionis}, {and} \bibinfo{person}{Prof
  Faujdar~Ram}.} \bibinfo{year}{2013}\natexlab{}.
\newblock \showarticletitle{Explaining the rural-urban gap in infant mortality
  in {India}}.
\newblock \bibinfo{journal}{\emph{Demographic Research}}  \bibinfo{volume}{29}
  (\bibinfo{year}{2013}), \bibinfo{pages}{473--506}.
\newblock


\bibitem[\protect\citeauthoryear{Salganik}{Salganik}{2019}]%
        {salganik2019bit}
\bibfield{author}{\bibinfo{person}{Matthew~J Salganik}.}
  \bibinfo{year}{2019}\natexlab{}.
\newblock \bibinfo{booktitle}{\emph{Bit by bit: social research in the digital
  age}}.
\newblock \bibinfo{publisher}{Princeton University Press}.
\newblock


\bibitem[\protect\citeauthoryear{Sen, Floeck, Weller, Weiss, and Wagner}{Sen
  et~al\mbox{.}}{2019}]%
        {sen2019total}
\bibfield{author}{\bibinfo{person}{Indira Sen}, \bibinfo{person}{Fabian
  Floeck}, \bibinfo{person}{Katrin Weller}, \bibinfo{person}{Bernd Weiss},
  {and} \bibinfo{person}{Claudia Wagner}.} \bibinfo{year}{2019}\natexlab{}.
\newblock \showarticletitle{A Total Error Framework for Digital Traces of
  Humans}.
\newblock \bibinfo{journal}{\emph{arXiv preprint arXiv:1907.08228}}
  (\bibinfo{year}{2019}).
\newblock


\bibitem[\protect\citeauthoryear{Sicular, Ximing, Gustafsson, and Shi}{Sicular
  et~al\mbox{.}}{2007}]%
        {sicular2007urban}
\bibfield{author}{\bibinfo{person}{Terry Sicular}, \bibinfo{person}{Yue
  Ximing}, \bibinfo{person}{Bj{\"o}rn Gustafsson}, {and} \bibinfo{person}{Li
  Shi}.} \bibinfo{year}{2007}\natexlab{}.
\newblock \showarticletitle{The urban--rural income gap and inequality in
  {China}}.
\newblock \bibinfo{journal}{\emph{Review of Income and Wealth}}
  \bibinfo{volume}{53}, \bibinfo{number}{1} (\bibinfo{year}{2007}),
  \bibinfo{pages}{93--126}.
\newblock


\bibitem[\protect\citeauthoryear{Sivak and Smirnov}{Sivak and Smirnov}{2019}]%
        {Sivak2039}
\bibfield{author}{\bibinfo{person}{Elizaveta Sivak} {and} \bibinfo{person}{Ivan
  Smirnov}.} \bibinfo{year}{2019}\natexlab{}.
\newblock \showarticletitle{Parents mention sons more often than daughters on
  social media}.
\newblock \bibinfo{journal}{\emph{Proceedings of the National Academy of
  Sciences}} \bibinfo{volume}{116}, \bibinfo{number}{6} (\bibinfo{year}{2019}),
  \bibinfo{pages}{2039--2041}.
\newblock
\urldef\tempurl%
\url{https://doi.org/10.1073/pnas.1804996116}
\showDOI{\tempurl}
\showeprint{https://www.pnas.org/content/116/6/2039.full.pdf}


\bibitem[\protect\citeauthoryear{Spyratos, Vespe, Natale, Weber, Zagheni, and
  Rango}{Spyratos et~al\mbox{.}}{2018}]%
        {spyratos2018migration}
\bibfield{author}{\bibinfo{person}{S Spyratos}, \bibinfo{person}{M Vespe},
  \bibinfo{person}{F Natale}, \bibinfo{person}{I Weber}, \bibinfo{person}{E
  Zagheni}, {and} \bibinfo{person}{M Rango}.} \bibinfo{year}{2018}\natexlab{}.
\newblock \showarticletitle{Migration Data using Social Media}.
\newblock \bibinfo{journal}{\emph{JRC Science Hub}} (\bibinfo{year}{2018}).
\newblock


\bibitem[\protect\citeauthoryear{Thorvaldsen}{Thorvaldsen}{[n.d.]}]%
        {thorvaldsenhandbook}
\bibfield{author}{\bibinfo{person}{G Thorvaldsen}.}
  \bibinfo{year}{[n.d.]}\natexlab{}.
\newblock \bibinfo{title}{Handbook of international historical microdata for
  population research}.
\newblock
\newblock


\bibitem[\protect\citeauthoryear{Tufekci}{Tufekci}{2014}]%
        {tufekci2014big}
\bibfield{author}{\bibinfo{person}{Zeynep Tufekci}.}
  \bibinfo{year}{2014}\natexlab{}.
\newblock \showarticletitle{Big questions for social media big data:
  Representativeness, validity and other methodological pitfalls}. In
  \bibinfo{booktitle}{\emph{Eighth International AAAI Conference on Weblogs and
  Social Media}}.
\newblock


\bibitem[\protect\citeauthoryear{Wicherts, Veldkamp, Augusteijn, Bakker,
  Van~Aert, and Van~Assen}{Wicherts et~al\mbox{.}}{2016}]%
        {wicherts2016degrees}
\bibfield{author}{\bibinfo{person}{Jelte~M Wicherts},
  \bibinfo{person}{Coosje~LS Veldkamp}, \bibinfo{person}{Hilde~EM Augusteijn},
  \bibinfo{person}{Marjan Bakker}, \bibinfo{person}{Robbie Van~Aert}, {and}
  \bibinfo{person}{Marcel~ALM Van~Assen}.} \bibinfo{year}{2016}\natexlab{}.
\newblock \showarticletitle{Degrees of freedom in planning, running, analyzing,
  and reporting psychological studies: A checklist to avoid p-hacking}.
\newblock \bibinfo{journal}{\emph{Frontiers in psychology}}
  \bibinfo{volume}{7} (\bibinfo{year}{2016}), \bibinfo{pages}{1832}.
\newblock


\bibitem[\protect\citeauthoryear{Yildiz, Munson, Vitali, Tinati, Holland,
  et~al\mbox{.}}{Yildiz et~al\mbox{.}}{2017}]%
        {yildiz2017using}
\bibfield{author}{\bibinfo{person}{Dilek Yildiz}, \bibinfo{person}{Joanna
  Munson}, \bibinfo{person}{Agnese Vitali}, \bibinfo{person}{Ramine Tinati},
  \bibinfo{person}{Jennifer Holland}, {et~al\mbox{.}}}
  \bibinfo{year}{2017}\natexlab{}.
\newblock \showarticletitle{Using Twitter data for demographic research}.
\newblock \bibinfo{journal}{\emph{Demographic Research}}  \bibinfo{volume}{37}
  (\bibinfo{year}{2017}), \bibinfo{pages}{1477--1514}.
\newblock


\bibitem[\protect\citeauthoryear{Young}{Young}{2013}]%
        {young2013inequality}
\bibfield{author}{\bibinfo{person}{Alwyn Young}.}
  \bibinfo{year}{2013}\natexlab{}.
\newblock \showarticletitle{Inequality, the urban-rural gap, and migration}.
\newblock \bibinfo{journal}{\emph{The Quarterly Journal of Economics}}
  \bibinfo{volume}{128}, \bibinfo{number}{4} (\bibinfo{year}{2013}),
  \bibinfo{pages}{1727--1785}.
\newblock


\bibitem[\protect\citeauthoryear{Zagheni, Garimella, Weber,
  et~al\mbox{.}}{Zagheni et~al\mbox{.}}{2014}]%
        {zagheni2014inferring}
\bibfield{author}{\bibinfo{person}{Emilio Zagheni}, \bibinfo{person}{Venkata
  Rama~Kiran Garimella}, \bibinfo{person}{Ingmar Weber}, {et~al\mbox{.}}}
  \bibinfo{year}{2014}\natexlab{}.
\newblock \showarticletitle{Inferring international and internal migration
  patterns from twitter data}. In \bibinfo{booktitle}{\emph{Proceedings of the
  23\textsuperscript{rd} International Conference on World Wide Web}}. ACM,
  \bibinfo{pages}{439--444}.
\newblock


\bibitem[\protect\citeauthoryear{Zagheni and Weber}{Zagheni and Weber}{2015}]%
        {zagheni2015demographic}
\bibfield{author}{\bibinfo{person}{Emilio Zagheni} {and}
  \bibinfo{person}{Ingmar Weber}.} \bibinfo{year}{2015}\natexlab{}.
\newblock \showarticletitle{Demographic research with non-representative
  internet data}.
\newblock \bibinfo{journal}{\emph{International Journal of Manpower}}
  \bibinfo{volume}{36}, \bibinfo{number}{1} (\bibinfo{year}{2015}),
  \bibinfo{pages}{13--25}.
\newblock


\bibitem[\protect\citeauthoryear{Zagheni, Weber, Gummadi,
  et~al\mbox{.}}{Zagheni et~al\mbox{.}}{2017}]%
        {zagheni2017leveraging}
\bibfield{author}{\bibinfo{person}{Emilio Zagheni}, \bibinfo{person}{Ingmar
  Weber}, \bibinfo{person}{Krishna Gummadi}, {et~al\mbox{.}}}
  \bibinfo{year}{2017}\natexlab{}.
\newblock \showarticletitle{Leveraging Facebook{\textquoteright}s advertising
  platform to monitor stocks of migrants}.
\newblock \bibinfo{journal}{\emph{Population and Development Review}}
  \bibinfo{volume}{43}, \bibinfo{number}{4} (\bibinfo{year}{2017}),
  \bibinfo{pages}{721--734}.
\newblock


\end{thebibliography}

\end{document}